\documentclass[nonacm,acmsmall, screen]{acmart} 

\AtBeginDocument{%
  }

\usepackage{amsmath,amsfonts}
\usepackage{graphicx}
\usepackage{ragged2e}
\usepackage{bm}
\usepackage{textcomp}
\usepackage{xcolor}
\usepackage{booktabs}   
\usepackage{listings}
\usepackage{color}
\usepackage{pifont}
\usepackage{wrapfig}
\usepackage{soul}
\usepackage[linesnumbered,lined,boxed,ruled,vlined]{algorithm2e}
\usepackage{mathtools}
\usepackage{hyperref}
\usepackage{tcolorbox}
\usepackage{longtable}
\usepackage{multirow}
\usepackage{balance}
\usepackage[switch]{lineno}

\newcommand{\REM}[1]{}

\newcommand{\sma}[1]{\textrm{\color{black} #1}}

\newcommand{\entp}{\textsf{EnTP}\xspace}
\newcommand{\hansie}{\textsf{Hansie}\xspace}

\newcommand{\colosseum}{\textsf{Colosseum}\xspace}

\newcommand{\shortpara}[1]{\vspace{2mm}\noindent\textbf{#1}}

\newcommand{\shouvick}[1]{\textcolor{blue}{{\it [Shouvick says: #1]}}}
\newcommand{\mycircled}[1]{\raisebox{0.3pt}{\textcircled{\raisebox{-0.5pt}{\fontsize{8}{8}\selectfont{\textsf{#1}}}}}}

\definecolor{brass}{rgb}{0.71, 0.65, 0.26}
\definecolor{indiagreen}{rgb}{0.07, 0.53, 0.03}
\definecolor{darkorange}{rgb}{1.0, 0.55, 0.0}
\definecolor{navyblue}{rgb}{0.36, 0.54, 0.66}
\definecolor{amber}{rgb}{1.0, 0.75, 0.0}

\newcommand{\smr}[1]{\textrm{\color{black} #1}}

\renewcommand\fbox{\fcolorbox{lightgray}{white}}
\newcommand{\CodeIn}[1]{{\small{\texttt{#1}}}}

\newcommand{\MyComment}[1]{}

\newcommand{\Processor}{AMD Ryzen 5 3550H CPU @ 2.10GHz}
\newcommand{\NumCPUs}{8}
\newcommand{\NumCores}{4}
\newcommand{\RAMCapacity}{8}
\newcommand{\HardDiskCapacity}{512}
\newcommand{\OSVersion}{Ubuntu 18.04}
\newcommand{\KernelVersion}{5.3.0-51-generic}
\newcommand{\GCCVersion}{5.5.0}
\newcommand{\LLVMVersion}{3.9.0}
\newcommand{\BashVersion}{4.4.20}

\begin{document}
\title{On Rank Aggregating Test Prioritizations}

\author{Shouvick~Mondal}
\affiliation{%
  \institution{Software Engineering and Testing (SET) Group, Indian Institute of Technology Gandhinagar}
  \city{Palaj}
  \state{Gujarat}
  \country{India}}
\email{shouvick.mondal@iitgn.ac.in}
\orcid{0000-0002-0703-8728}

\author{Tse-Hsun~(Peter)~Chen}
\affiliation{%
  \institution{Software PErformance, Analysis, and Reliability (SPEAR) lab, Concordia University}
  \city{Montreal}
  \state{Quebec}
  \country{Canada}}
\email{peterc@encs.concordia.ca}
\orcid{0000-0003-4027-0905}

\renewcommand{\shortauthors}{Mondal et al.}

\begin{abstract}
Test case prioritization (TCP) has been an effective strategy to optimize regression testing. Traditionally, test cases are ordered based on some heuristic and rerun against the version under test with the goal of yielding a high failure throughput. Almost four decades of TCP research has seen extensive contributions in the light of individual prioritization strategies. However, test case prioritization via preference aggregation has largely been unexplored. We envision this methodology as an opportunity to obtain robust prioritizations by consolidating multiple standalone ranked lists, i.e., performing a consensus. In this work, we propose Ensemble Test Prioritization (\entp) as a three stage pipeline involving: (i) ensemble selection, (ii) rank aggregation, and (iii) test case execution. We evaluate \entp on 20 open-source C projects {from the Software-artifact Infrastructure Repository and GitHub} (totaling: 694,512 SLOC, 280 versions, and 69,305 system level test-cases). We employ an ensemble of 16 standalone prioritization plans, four of which are imposed due to respective state-of-the-art approaches. We build \entp on the foundations of \hansie, an existing framework on consensus prioritization and show that \entp's diversity based ensemble selection budget of top-75\% followed by rank aggregation can outperform \hansie, and the employed standalone prioritization approaches.
\end{abstract}

%

\maketitle



\section{Introduction}\label{sec:intro}
\begin{figure}
    \centering
    \fbox{\hspace{37pt}\includegraphics[width=0.565\linewidth]{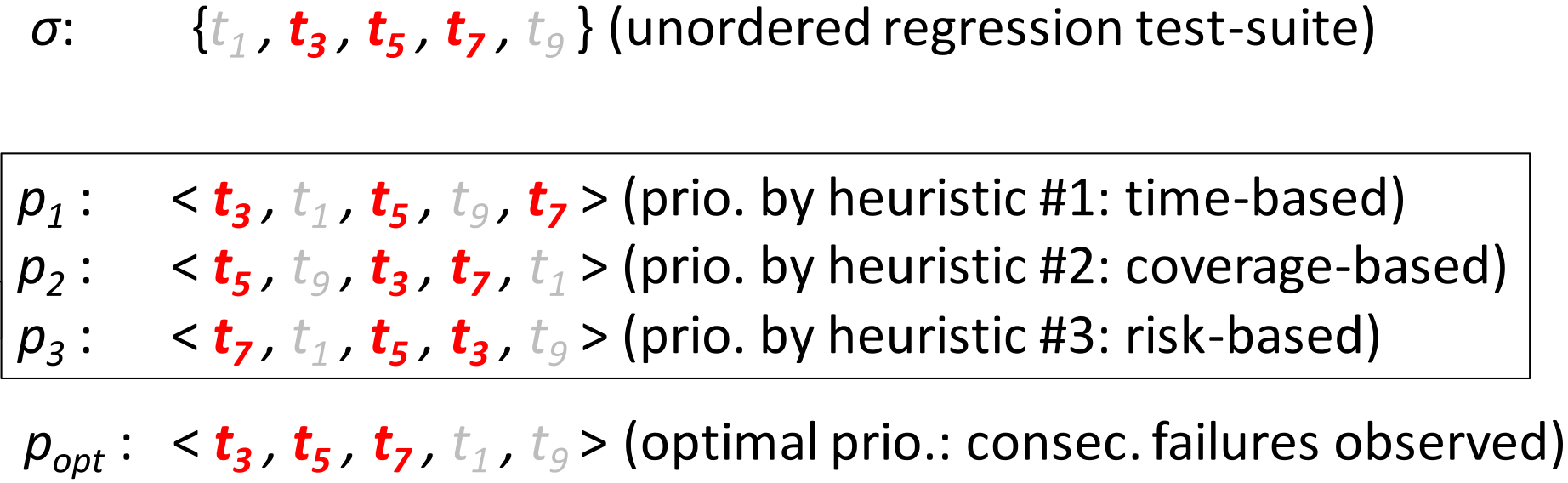}}
    \fbox{\includegraphics[width=0.675\linewidth]{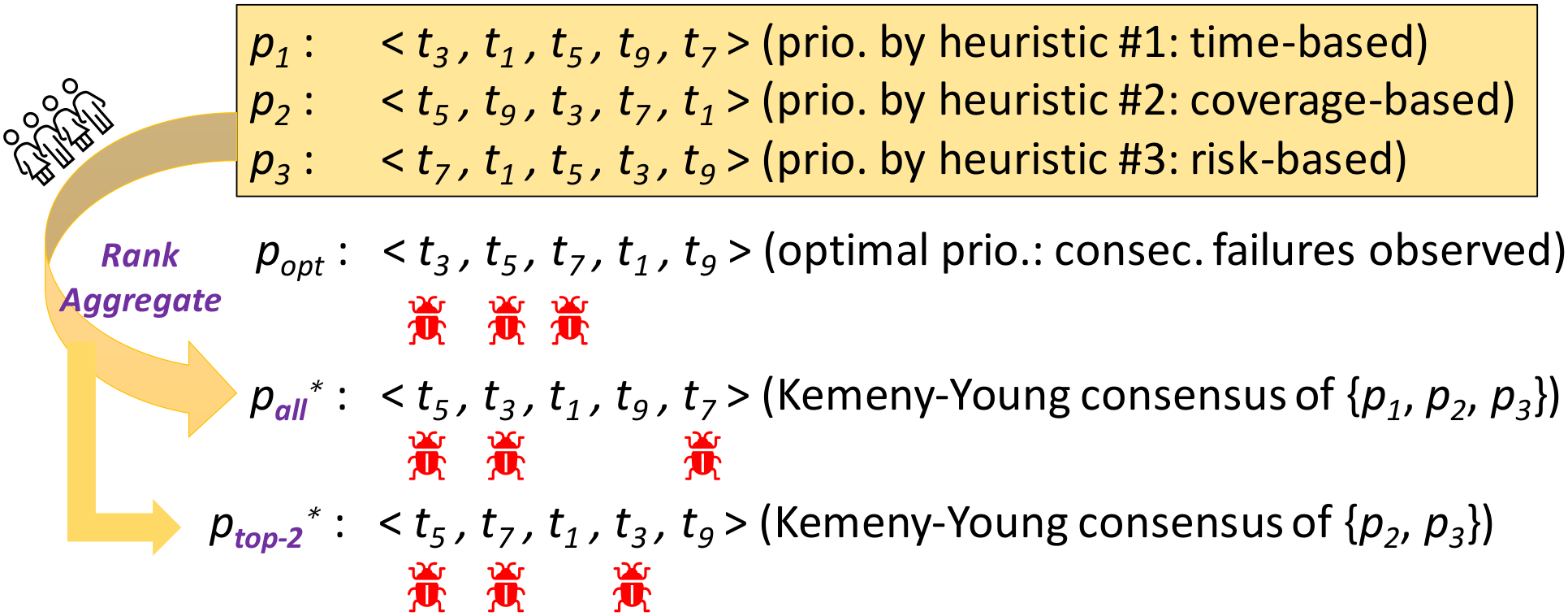}}
    \caption{Conventional prioritization ranking (top), versus consensus prioritization by rank aggregation (bottom). Failed test cases are marked in \textbf{\color{red}red}.}
    \label{fig:example}
\end{figure}

Software regression testing has been an extensively researched topic in software quality assurance. 
Previous studies have reported that regression testing can consume as much as 80\% of the overall testing budget and account for 50\% of software maintenance costs~\cite{Bertolino_2007,Harrold_2009}. Hence, the sequence in which regression test cases are executed and failures are observed, determines the rate at which regression faults are detected. This reduces the turn around time to report feedback on test failures. Hence, the order of test execution is critical to regression testing. The scale~\cite{yaraghi2021tcp} of today's system under test (SUT) coupled with non-uniformity of test cases (i.e., execution costs: a few seconds to several hours) poses a unique challenge. This problem is addressed by test case prioritization~\cite{yaraghi2021tcp,10.1002/stvr.1572,Yoo_2012_survey}, an effective optimization tool for software regression testing. Although mostly, a test case (when executed) has a binary pass/fail verdict, the optimization of benefit that yields from executing multiple test cases in sequence happens to be \textit{early} and \textit{closely spread} test failure observation rate. For instance, with the code change (Line~\ref{line:v1}: $n^2\rightarrow{2n}$) in Figure~\ref{fig:source1} with \textit{five} test cases ($\{t_1,t_3,t_5,t_7,t_9\}$: change-traversing + originally intended for \texttt{source.c}) and \textit{three} failures ($\{{\color{red}t_3},{\color{red}t_5},{\color{red}t_7}\}$: \textit{post execution} on \texttt{source-v1.c}), the optimized execution scenario corresponds to vector $\langle1,1,1,0,0\rangle$, where `1' denotes test case failure. This is shown in Figure~\ref{fig:example} (top) as $p_{\textnormal{\textit{opt}}}: \langle{{\color{red}t_3},{\color{red}t_5},{\color{red}t_7},{\color{gray}t_1},{\color{gray}t_9}}\rangle$. All other scenarios (heuristics) may turn out to be sub-optimal, say for illustration: (i) $p_1$ (time based~\cite{Chen_2018}) yields vector $\langle1,0,1,0,1\rangle$, $p_2$: (code-coverage based~\cite{Jasz_2012}) yields vector $\langle1,0,1,1,0\rangle$, and $p_3$: (risk based~\cite{risk_prio}) yields vector $\langle1,0,1,1,0\rangle$. 

Unfortunately, the optimal sequence $p_{\textnormal{\textit{opt}}}$ is hard to obtain statically as it is very difficult to know \textit{apriori} whether a test case would fail when executed against a new version of the software. To cope with this difficulty and obtain a reasonable optimization, we resort to prioritization heuristics that strive to achieve a close approximation by constructing a sequence from the set of regression test cases to be executed. Post execution, the sequence of the associated verdicts are evaluated in terms of two popular metrics: (i) Average Percentage of Faults Detected (APFD)~\cite{APFD_2001,APFD_2002}, and (ii) Cost-cognizant APFD (APFD$_c$)~\cite{APFD_c} that takes into account execution costs of test cases. The values of these metrics lie in the interval [0,1] with values close to `1' denoting more effective test case prioritizations. 

\begin{figure}[h]
\centering
\begin{tabular}{l@{\hskip 0.06\columnwidth}l}
{\begin{lstlisting}
// source.c	
#include<stdio.h>
int f(int n)
{
	int result=-1;
	
	if(n % 2 == 0)
		result = n * n;
	else
		result = n + 3 * n;
	
	return(result);
}
int main()
{
	int n;
	scanf("%d", &n);
	printf("%d\n",f(n));
	return(0);
}
\end{lstlisting}}
&
{\begin{lstlisting}
// source-v1.c
#include<stdio.h>
int f(int n)
{
	int result=-1;

	if(n % 2 == 0)
		|{\color{indiagreen} result = n * 2;}| //change|\label{line:v1}|
	else
		result= n + 3 * n;

	return(result);
}
int main()
{
	int n;
	scanf("%d", &n);
	printf("%d\n",f(n));
	return(0);
}
\end{lstlisting}}%
\end{tabular}
\begin{center}
\small
$\Sigma$: $\{t_1,t_2,t_3,...,t_{10}\}$ (original test-suite)\\
$\{t_1,...,t_{10}\}$: (contains randomly generated integer in $[0..99]$)\\
$\{0,4,16,50,2\}$: (integers present in $\{t_1,t_3,t_5,t_7,t_9\}$)\\
All other test cases $\in{\Sigma}$ contain odd numbers in $[0..99]$
\end{center}
\caption{Original C source (\textit{source.c} from~\cite{hansie_2020}) and its modified revision (\textit{source-v1.c}). Test-suite description appears below the code.}
\label{fig:source1}
\end{figure}

Since there is no one optimal heuristic that faithfully produces an optimal sequence of test failures for any arbitrary code change and software under test, a plethora of heuristics had been designed since the last four decades~\cite{Yoo_2012_survey,Kazmi_2017,Khatibsyarbini_2018} and continues to be an active topic nowadays~\cite{BTCP_2022,Peng_2020,Joo_ICST_2020_prop,Huang_2020,Vescan_2020,mahtab_2019,col_tse_2021} indicating that the research community has not yet achieved an optimal enough solution. 
A recent work by Mondal and Nasre~\cite{hansie_2021} proposed \hansie, the first-of-its-kind framework that firstly computes multiple permutations of the unordered regression test-suite. Then, instead of executing any standalone permutation, \hansie constructs a consensus prioritization sequence that on average outperforms the standalones. However, \hansie rank-aggregates the entire ensemble of standalone permutations which may turn out to be ineffective as consensus is worth taking \textit{only} when individual permutations are diverse. The metaphor we use here is: \textit{there are as many opinions as there are experts, but it is worth taking the consensus of not all but some diverse opinions rather than similar ones}. Hence, there is scope for further improving the effectiveness of consensus prioritization. 

In this paper, we propose ensemble test prioritization (\entp) to achieve this. The comparison of \entp against traditional consensus test case prioritization (\hansie) is illustrated in Figure~\ref{fig:example} (bottom) where the failure vectors are: \hansie ($\langle1,1,0,0,1\rangle$ with ${p_{\textnormal{\textit{all}}}}^*$) versus \entp ($\langle1,1,0,1,0\rangle$ with ${p_{\textnormal{\textit{top-2}}}}^*$), the latter one being closer to $p_{\textnormal{\textit{opt}}}$ by gaining a position early for the last failure.
\entp enriches \hansie along two directions: (i) by including and repeating more effective standalone heuristics~\cite{col_tse_2021,APFD_c,Epitropakis_2015} than was originally considered, (ii) by performing a budgeted rank-aggregation by considering only the top-$k\%$ diverse~\cite{Kendall_1938,Kendall_1948} permutations. Compared to non-consensus based TCP approaches, our approach in \entp is orthogonal to the existing literature~\cite{col_tse_2021,Epitropakis_2015,Elbaum_2014,Mostafa_2017,Wang_2017_2,FAST_2018,Jiang_2009,STVR_hybrid_2019,Walcott_2006,Busjaeger_2016,Chi_2020,Joo_ICST_2020_prop}. The major difference lies in the way the final prioritization sequence is constructed: \entp follows social choice theoretic principles~\cite{Truchon_1998,social_choice_theory}, whereas all the aforementioned works are domain specific to software testing. In this work, we set out to empirically investigate on whether consensus prioritization powered by diversity based ensemble selection yields benefit.

The main contributions of this paper are as follows.

\begin{itemize}
	\item{We propose ensemble test prioritization (\entp) as a three stage pipeline of diversity based ensemble selection, consensus rank aggregation, and prioritized test execution.}
        \item{We evaluate \entp on a test bed of 20 open-source C projects (totaling: 694,512 SLOC, 280 versions, and 69,305 test-cases having high load/cost imbalance), and empirically show that \entp with an aggregation budget of top-75\% outperforms state-of-the-art approaches.}
	\item{We release the artifacts publicly for data availability and independent reproduction of experimental results~\cite{replicationPackage}.}
\end{itemize}

\section{Our approach: \entp}\label{sec:approach}
The general idea behind our approach is to obtain a potentially robust prioritization of test cases by consolidating a collection of heuristically independent test case orderings. 
\begin{figure}[h]
    \centering
    \includegraphics[width=0.38\linewidth]{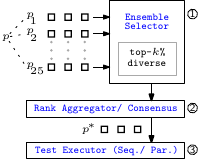}
    \caption{Workflow of \entp.}
    \label{fig:workflow}
\end{figure}
To achieve this, we leverage techniques from computational social choice~\cite{Truchon_1998,social_choice_theory}, which views agreement among decision-makers as an objective function. The goal in this setting is to maximize social welfare -- the maximization of the agreements in a collective manner. The process involves multiple voters (decision-makers), and multiple candidates (objects of interest). The decision of a voter imposes an ordering or a preferences list of the objects. Namely, two voters, $v_x$ and $v_y$, agree on a pair of objects $(i,j)$ if and only if the relative ordering between the objects has the same preference ($i$ before $j$, or $j$ before $i$) for both $v_x$ and $v_y$. Thus, the collective agreement over all pairs of objects and across all voters demands the computation of an aggregated preference ordering that maximizes the sum total of individual agreements. In rank aggregation, this corresponds to computing \textit{Kendall-tau (KT) rank distance}~\cite{Kendall_1938,Kendall_1948,Kemeny_1959} between permutations. In this work, we leverage the same concept for test prioritization where: (i) voters correspond to standalone prioritization heuristics, (ii) candidates correspond to regression test cases, and (iii) permutations correspond to prioritization plans (yet to be executed test cases). Below, we describe the details of \entp.

\shortpara{Kendall-tau rank distance.} The $KT$-distance quantifies the total number of pairwise disagreements (inversions) between two permutations. Disagreement for a pair $(i,j)$\footnote{In the context of our work, these denote pair of test case identifiers.} with $(i<j)$ is one $(100\%)$ if the order of their appearance in two prioritizations (ranked lists) is inverted. Otherwise, the disagreement is zero $(0\%)$. This can be formulated as follows~\cite{hansie_2020}:
\begin{eqnarray}
&KT(a,b)=|\{(i,j): (i<j) \wedge \textnormal{($C_1\vee{C_2}$)}\}|\nonumber\\
&C_1: \{\rho(a_i)<\rho(a_j)\} \wedge \{\rho(b_i)>\rho(b_j)\}\nonumber\\
&C_2: \{\rho(a_i)>\rho(a_j)\} \wedge \{\rho(b_i)<\rho(b_j)\}
\label{eqn:KT-eqn}
\end{eqnarray}

\noindent where $\rho(a_i)$ denotes the rank of element $i$ in the permutation denoted by $a$. A lower $\rho$ value indicates early occurrence of $i$ in $a$. $KT(a,b)$ is $0$ if $a$ and $b$ are equal. As an example, for the set $\{t_1,t_3,t_5,t_7,t_9\}$, the $KT$-distance between $\langle t_3,t_1,t_5,t_9,t_7\rangle$ and $\langle t_5,t_9,t_3,t_7,t_1\rangle$ is $5$ because of five inverted pairs: $(t_1, t_5)$, $(t_1, t_7)$, $(t_1, t_9)$, $(t_3, t_5)$, and $(t_3, t_9)$.
\bgroup

\begin{table}[h]
\caption{Standalone prioritizations constituting the ensemble}
		{\fontsize{7.5}{7.5}\selectfont
            \begin{minipage}{\linewidth}
            \begin{center}
			\begin{tabular}{@{}l|l|l|l|r@{}} 
                    
				\toprule
				{\textbf{Family}}&{\textbf{Strategy} \fbox{\fontsize{6.5}{6.5}\selectfont\textsf{label}}} & {\textbf{Heuristic}} & {\textbf{Type}} & {\textbf{Ref.}}\\
				\midrule
				{$\Delta$-coverage} & {relevance \fbox{\fontsize{6.5}{6.5}\selectfont\textsf{rel}}} & {largest affected ($\Delta$)} & {single} & {~\cite{mahtab_2019}}\\
				{} & {(greedy total)} & {coverage ($cov\left(t\right)$) first} & {} & {~\cite{APFD_2001}}\\
				{} &{} & \fbox{$rel(t)= \frac{|cov(t)\cap\Delta|}{|\Delta|}$} & {} & {~\cite{APFD_2002}}\\
				{} &{} & {cover \textit{affected} code} & {} & {}\\
				{} &{} & {as \textit{much} as possible} & {} & {}\\
				\midrule
				{$\Delta$-coverage} &{confinedness \fbox{\fontsize{6.5}{6.5}\selectfont\textsf{con}}} & {largest undeviated} & {single} & {~\cite{mahtab_2019}}\\
				{} &{} & {(confined) coverage first} & {} & {}\\
				{} &{} & \fbox{$con(t)=|cov(t)-\Delta|^{-1}$} & {} & {}\\
				{} &{} & {cover \textit{unaffected} code} & {} & {}\\
				{} &{} & {as \textit{less} as possible} & {} & {}\\
				\midrule
				{exec. time} &{cost-only \fbox{\fontsize{6.5}{6.5}\selectfont\textsf{cost}} 4$\times$} & {shortest execution} & {single} & {~\cite{Chen_2018}}\\
				{} &{} & {time first} & {} & {}\\
				\midrule
				{$\Delta$-coverage} &{greedy \fbox{\fontsize{6.5}{6.5}\selectfont\textsf{ga}}} & {largest remaining} & {single} & {~\cite{APFD_2001}}\\
                    {} &{additional} & {coverage first} & {} & {~\cite{APFD_2002}}\\
                    \midrule
    {$\Delta$-coverage \&} &{cost-cognizant \fbox{\fontsize{6.5}{6.5}\selectfont\textsf{costga}}} & {largest remaining coverage} & {single} & {~\cite{Epitropakis_2015}}\\
				{exec. time} &{greedy additional} & {per time unit first} & {} & {}\\
				\midrule
				{$\Delta$-coverage} &{\fbox{{\fontsize{6.5}{6.5}\selectfont\textsf{relcon\_10\_90}}}} & {$0.1*rel(t)+0.9*con(t)$\footnote{Largest weighted-score first.}} & {hybrid} & {~\cite{hansie_2020}}\\
				{} &{\fbox{{\fontsize{6.5}{6.5}\selectfont\textsf{relcon\_20\_80}}}} & {$0.2*rel(t)+0.8*con(t)$} & {hybrid} & {~\cite{hansie_2020}}\\
				{} &{\fbox{{\fontsize{6.5}{6.5}\selectfont\textsf{relcon\_30\_70}}}} & {$0.3*rel(t)+0.7*con(t)$} & {hybrid} & {~\cite{hansie_2020}}\\
				{} &{\fbox{{\fontsize{6.5}{6.5}\selectfont\textsf{relcon\_40\_60}}}} & {$0.4*rel(t)+0.6*con(t)$} & {hybrid} & {~\cite{hansie_2020}}\\
				\midrule
				{$\Delta$-coverage} &{\fbox{{\fontsize{6.5}{6.5}\selectfont\textsf{relcon\_50\_50}}}} & {$0.5*rel(t)+0.5*con(t)$} & {hybrid} & {~\cite{mahtab_2019}}\\
				{} &{(\textit{relevance and}} & {} & {} & {}\\
				{} &{\textit{confinedness})} & {} & {} & {}\\
				\midrule
				{$\Delta$-coverage} &{\fbox{{\fontsize{6.5}{6.5}\selectfont\textsf{relcon\_60\_40}}}} & {$0.6*rel(t)+0.4*con(t)$} & {hybrid} & {~\cite{hansie_2020}}\\
				{} &{\fbox{{\fontsize{6.5}{6.5}\selectfont\textsf{relcon\_70\_30}}}} & {$0.7*rel(t)+0.3*con(t)$} & {hybrid} & {~\cite{hansie_2020}}\\
				{} &{\fbox{{\fontsize{6.5}{6.5}\selectfont\textsf{relcon\_80\_20}}}} & {$0.8*rel(t)+0.2*con(t)$} & {hybrid} & {~\cite{hansie_2020}}\\
				{} &{\fbox{{\fontsize{6.5}{6.5}\selectfont\textsf{relcon\_90\_10}}}} & {$0.9*rel(t)+0.1*con(t)$} & {hybrid} & {~\cite{hansie_2020}}\\
                 \midrule
			{$\Delta$-displacement} &{colosseum {\fbox{{\fontsize{6.5}{6.5}\selectfont\textsf{coluw}}}}} & {shortest delta propagation} & {single} & {~\cite{col_tse_2021}}\\
                    {} &{(\textit{unweighted})} & {displacement path first} & {} & {}\\
                    {} & {} & {} & {} & {}\\
                    {} &{colosseum {\fbox{{\fontsize{6.5}{6.5}\selectfont\textsf{colw}}}} 7$\times$} & {\#instructions in CFG basic} & {single} & {~\cite{col_tse_2021}}\\
                    {} &{(\textit{weighted})} & {blocks define resp. weights} & {} & {}\\
				\bottomrule
			\end{tabular}
            \begin{center}
            \textit{state-of-the-art}:~\{\texttt{rel}, \texttt{cost}, \texttt{ga}, \texttt{costga}\}
            \end{center}
            \end{center}
            \end{minipage}
		}
		\label{tab:strategies}  
 \fbox{\fontsize{7.5}{7.5}\selectfont \textbf{Ensemble size = 25 rankings} (17 w/o + 8 w/ $\Delta$-displacement)}
\end{table}
\egroup

\noindent{\bf The design of \entp}.
\entp is composed of a three-stage pipeline as depicted in Figure~\ref{fig:workflow}. The input ($p$) is the unordered sequence of regression test cases, selected based on the changes ($\Delta=\{\delta_1,...,\delta_m\}$) made to the new version of the software, henceforth referred to as software under test (SUT). Subsequently, as a pre-processing step, $p$ is independently subjected to 16 different standalone prioritizations what were proposed in prior studies, each of which returns a different permutation of $p$ based on the associated heuristic. Table~\ref{tab:strategies} presents a brief description of each heuristic in a separate row. The column \textbf{Family} specifies the generic class of a prioritization strategy. There are four families of strategies considered: (i) $\Delta$-coverage: which takes into account only the set of affected basic blocks in the program, (ii) exec. time: which considers only the cost of running a test case, (iii) $\Delta$-coverage \& exec. time, and (iv) $\Delta$-displacement: which approximates the probability of test case to combat failed-error-propagation by considering the propagation of the $\Delta$ along the loop-free in-order execution trace (only first occurrence per basic-block) of a test case (old version\footnote{This is necessary as we cannot execute the test case on the new version while constructing the prioritization plan which occurs prior to exercising the SUT.}). The column \textbf{Strategy} presents the name and label (identifier) of the heuristic. The descriptions of these heuristics (along with priority computation for a test case $t$) are briefly mentioned in the next column (\textbf{Heuristic}). The Delta ($\Delta$)-displacement based heuristic (\colosseum)~\cite{col_tse_2021}, considers stale (older version) basic block traces of each regression test-case's execution path. The traces are linearized to paths by keeping only the first visit of a basic block. \colosseum associates three parameters with each of these paths: the displacement $\alpha$ of the first delta from the starting basic-block, the displacement $\gamma$ of the last delta from the terminating basic-block, and the average displacement $\beta$ within all the intermediate basic-blocks. The intuition is that regression test cases with shorter overall displacement are more likely to propagate the effects of the code changes to the program's observable outputs. The assumption~\cite{col_tse_2021} is that the likelihood of a test-case detecting a regression fault depends on the probability of change propagation, which can potentially result in an error. Hence, by extension, delta displacement provides an estimate of failed error propagation~\cite{Androutsopoulos_2014_prop,oracle_jahangirova_2017}. The unweighted variant (\texttt{coluw}) considers all delta blocks alike, but the weighted variant (\texttt{colw}) considers \#instructions contained in a delta basic block as its \textit{weight}. The idea is that blocks with a higher number of instructions tend to mask local changes computationally, i.e., due to more interference it is less likely to propagate changes (intermediate error states) to successor basic-blocks and ultimately to the output. Empirically~\cite{col_tse_2021}, the weighted variant (\texttt{colw}) is more effective.

In Table~\ref{tab:strategies}, the column \textbf{Type} also mentions whether a heuristic is hybrid, i.e., linear combination of two different heuristics: relevance (\texttt{rel}) and confinedness (\texttt{con})~\cite{mahtab_2019,hansie_2020}. \entp's expanded ensemble (size 25) configuration over \hansie~\cite{hansie_2021} (size 13) is explained as follows.

\shortpara{Ensemble configuration.} Regarding the prioritizations constituting the ensemble, \hansie had 13 different strategies with an unbalanced distribution of heuristics: relevance-and-confinedness (\texttt{relcon}) based ($11\times$), (ii) cost-based (\texttt{cost}), (iii) greedy additional (\texttt{ga}). To strike a balance and further improve upon \hansie, \entp \textit{expands the ensemble based on empirical inspection for the best result} as follows: (i) assigns a weight of 4 to \texttt{cost} by repeating its permutation 4 times, (ii) adds a new heuristic (\texttt{costga}) which is state-of-the-art cost-aware white-box approach, and (iii) incorporates two variants (\texttt{coluw}, \texttt{colw}) of \colosseum. \entp repeats the permutations for the weighted variant (\texttt{colw}) a total of $7\times$ for two reasons: \textit{firstly}, \texttt{colw} empirically~\cite{col_tse_2021} outperforms \texttt{coluw}, \textit{secondly}, the ensemble is biased only towards $\Delta$-coverage based heuristics which do not consider $\Delta$-displacement. The final ensemble configuration (rankings: $E=\{p_1,...,p_{25}\}$ to be aggregated) is as follows.
\begin{center}
\fbox{\fontsize{8}{8}\selectfont \textbf{Ensemble size = 25 rankings} (17 w/o + 8 w/ $\Delta$-displacement)}
\end{center}
\noindent
The rest of the workflow of \entp (Figure~\ref{fig:workflow}) is explained in Sections~\ref{secd:diversity},~\ref{sec:consensus}, and~\ref{sec:execution}.

\subsection{Ensemble selection (stage \mycircled{1} of 3)}\label{secd:diversity}

The entire ensemble ($E$) of 25 rankings is fed to the \texttt{Ensemble Selector} which selects only the top-$k\%$ diverse permutations in terms of $KT$-distance. The budget of $k\%$ is a user input for stage \mycircled{1}. The selection process is as follows.

\begin{itemize}
    \item[]{\textbf{Step 1.} For each $p_i\in{E}$, compute the diversity $Div(p_i,E)$ as the sum of the $KT$-distances (equation~\ref{eqn:KT-eqn}) between $p_i$ and all other $p_{j\neq{i}}\in{E}$.}
    \item[]{\textbf{Step 2.} Sort $p_i$s in decreasing order of $Div()$ values.}
    \item[]{\textbf{Step 3.} Select top-($k\%$ of $|E|$) $p_i$s.}
\end{itemize}

For our running example, the selection process yields top-2 diverse permutations. The computations are as follows.

{\fontsize{8.5}{8.5}\selectfont
\begin{eqnarray}
\begin{aligned}
&E=\{p_1:\langle t_3,t_1,t_5,t_9,t_7\rangle,p_2:\langle t_5,t_9,t_3,t_7,t_1\rangle,p_3:\langle t_7,t_1,t_5,t_3,t_9\rangle\}\nonumber\\
&Div(p_1,E)=KT(p_1,p_2)+KT(p_1,p_3)=5+6=11~\textnormal{(rank: 3)}\nonumber\\
&Div(p_2,E)=KT(p_2,p_1)+KT(p_2,p_3)=5+7=12~\textnormal{(rank: 2)}\nonumber\\
&Div(p_3,E)=KT(p_3,p_1)+KT(p_3,p_2)=6+7=13~\textnormal{(rank: 1)}\nonumber\\
&\textnormal{\textit{top}-2 diverse prioritizations in \textit{E}:}~\{p_3,p_2\}\nonumber
\end{aligned}
\end{eqnarray}
}

The ensemble subset (with diversity) thus obtained is passed on to stage \mycircled{2} where we leverage social choice theoretic principles to perform the rank-aggregation to obtain the consensus prioritization plan of the regression test cases.

\subsection{Consensus prioritization (stage \mycircled{2} of 3)}\label{sec:consensus}

Given a profile of prioritizations $R=\{p_1,...,p_n\}$, the task of computing a consensus $p^*$ amounts to minimizing the sum of permutation distances (disagreements) of $p*$ form each of $p_i$'s in $R$. This can be mathematically stated as follows~\cite{hansie_2020}. Given a permutation distance metric $d(p_i,p_j)$ and permutations $R=\{p_1,p_2,...p_n\}$ as input, find an optimal permutation $p^*$ such that~\cite{Gattaca_2014}:
\begin{equation}
	p^*=\arg\min_{p}\sum\limits_{i=1}^{n}d(p_i,p)
\label{eqn:agg}
\end{equation}

We enforce input prioritizations $p_i$s to be strictly ordered such that no two test cases have the same rank. Otherwise, ties in input may lead to a tie in the final consensus as well. Even if ties are not present in the input, it is still possible for $p^*$ to have ties for two test cases $t_i$, and $t_j$. This scenario arises when then the number of preferences $|t_i\rightarrow{t_j}|=|t_j\rightarrow{t_i}|$ (aversions) in the profile $R$. It is due to this nature of ties in the consensus that a consensus test schedule may be respected by executing tied test cases within parallel execution windows (batches of isolated processes), and sequentially otherwise. The following subsections explain different ways to construct a consensus prioritization plan.

\subsubsection{Prioritization by Kemeny-Young consensus}
The Kemeny-Young (KY) consensus~\cite{Kemeny_1959,Young_1978,Dwork_2001,Gattaca_2014,Fagin_2003,Fagin_2004,Betzler_2008,Kumar_2010} rank aggregates a collection of preferences (ordering of candidates) imposed by multiple voters. The goal is to construct an consensus ordering such that the sum total of the Kendall-tau distances of the consensus from the individual ordering is minimal. The mathematical formulation of the consensus prioritization (\smr{equation}~\ref{eqn:agg}) deals with a general permutation distance function, $d$. For KY-consensus, we substitute $d$ by Kendall-tau ($KT$) distance. An equivalent formulation of the KY-consensus computes a sequence of candidates (test cases) that has the highest overall ranking score, wherein the score of a sequence (potential consensus prioritization) is the sum total of pairwise preference of test cases. If test case $t_i$ occurs before $t_j$ in the candidate consensus, the score of $t_i\rightarrow t_j$ (preferring $t_i$ to $t_j$) denotes how many standalone orderings support this preference. Continuing in this manner, the final consensus ordering is the one that has the most satisfying (to the extent possible) pairwise preferences. However, ties may appear for a test case pair $(t_i,t_j)$ in the majority preference, if the number of prioritizations preferring $t_i\rightarrow t_j$ exactly equals the count of aversions, i.e., preferring $t_j\rightarrow t_i$. 
For these situations, tie-breaking amounts to selecting an ordering from among those with the same overall score. In the final consensus prioritization, $t_i$ appears before $t_j$ if and only if the majority ($>50\%$) of the prioritizations prefer $t_i\rightarrow t_j$.

KY-consensus is NP-hard and is computationally expensive even for four~\cite{Dwork_2001} voters, i.e., standalone prioritization orderings. In this work, we follow the simplification as inspired by Mondal and Nasre~\cite{hansie_2020,hansie_2021}. We employ a quick approximation by occasionally leveraging randomization. This procedure is explained as follows. Given a profile of prioritizations, an initial approximation to the consensus is obtained by assigning to each test case the average (arithmetic mean) of the ranks due to individual orderings. This step is followed by further refinement of the consensus through $N$ independent iterations. Each iteration randomly shuffles a sub-sequence (of size $M$) of the initial approximation and re-evaluates its ranking score. This shuffling window of size $M$ slides along the entire sequence and the overall iterative procedure terminates on reaching a fixed point in the overall agreement score for a sequence. Several candidate consensus orderings may have identical maximum agreement scores. Such orderings are termed as Kemeny optimal orderings. Should this situation arise, the final consensus ordering is determined (as in~\cite{hansie_2020,hansie_2021}) by averaging ranks for every test case across all iterations producing optimal orderings. This ensures that in the final KY-consensus, no two test case pairs are ever rank-tied for a particular priority. 

The end-to-end time complexity of this approach is $O(N\times2^M)$~\cite{KY_C++}. The ranges of parameters in our empirical evaluation were similar to what was followed in \hansie by Mondal and Nasre~\cite{hansie_2020,hansie_2021}: $0\le{N}\le{200}$, and $1\le{M}\le{7}$.
For our running example with regression test-suite of five test cases: $\{t_1,t_3,t_5,t_7,t_9\}$ and the top-2 diverse prioritizations in $E$ (profile $R=\{p_2,p_3\}$) with $p_2=\langle t_5,t_9,t_3,t_7,t_1\rangle$, and $p_3=\langle t_7,t_1,t_5,t_3,t_9\rangle$, the \textit{KY-consensus} is $\langle t_5,t_7,t_1,t_3,t_9\rangle$.

\subsubsection{Prioritization by Borda-count method}\label{sec:borda}

The Kemeny-Young consensus is computationally expensive and should be used with small-sized regression test-suites. To alleviate the scalability issue a quick consensus approach is the Borda-count (BC) method~\cite{Borda,Dwork_2001,Gattaca_2014,Davenport_2004}. This method assigns to each test case $t$ the number of test cases behind it in each of the prioritizations in $R$. This score is referred to as the Borda-count score (defeat strength), and the consensus prioritization is obtained by arranging the test cases in terms of their decreasing Borda-count score. The overall time complexity of the Borda-count method turns out to be $O(mn^2+mn+n\log_2n)$~\cite{hansie_2020,hansie_2021}, with $m$ prioritizations, and $n$ test cases. For our running example with regression test-suite of five test cases: $\{t_1,t_3,t_5,t_7,t_9\}$ and the top-2 diverse prioritizations in $E$ (profile $R=\{p_2,p_3\}$) with $p_2=\langle t_5,t_9,t_3,t_7,t_1\rangle$, and $p_3=\langle t_7,t_1,t_5,t_3,t_9\rangle$, the BC-score is computed as follows:

{\fontsize{8.5}{8.5}\selectfont
\begin{eqnarray}
&\textnormal{BC-score$(t_1)=0+3=3$}\nonumber\\
&\textnormal{BC-score$(t_3)=2+1=3$}\nonumber\\
&\textnormal{BC-score$(t_5)=4+2=6$}\nonumber\\
&\textnormal{BC-score$(t_7)=1+4=5$}\nonumber\\
&\textnormal{BC-score$(t_9)=3+0=3$}\nonumber\\
&\textnormal{\textit{BC-consensus}=$\langle t_5,t_7,[t_1,t_3,t_9]\rangle$}\nonumber
\end{eqnarray}
}

\sma{For $t_1$ the BC-score from $p_2$ is $0$ as it defeats none. From $p_3$ the BC-score is $3$ as it defeats three test cases: \{$t_5,t_3,t_9\}$. Therefore, the total BC-score for $t_1$ in $R$ turns out to be $3$ $(0+3)$.} The final consensus ordering due to the Borda-count method contains ties for rank $3$ ($[t_1,t_3,t_9]$) with tied test cases enclosed in square brackets.

\subsubsection{Consensus approximation by mean and median}\label{sec:stats}

This category may be viewed as a family of quick-initial consensus approximations. This initial approximation is essentially the first step followed in the KY consensus. However, we do not further the iterative refinements but terminate the process prematurely with a crude approximation. The consensus operators employed are arithmetic mean (AM), geometric mean (GM), harmonic mean (HM), and median (med.). The time complexity for this category of consensus is $O(mn^2+mn+n\log_2n)$~\cite{hansie_2020,hansie_2021}, with $m$ prioritizations, and $n$ test cases. For our running example with $p_2=\langle t_5,t_9,t_3,t_7,t_1\rangle$, and $p_3=\langle t_7,t_1,t_5,t_3,t_9\rangle$, the AM-consensus is computed as follows:

{\fontsize{8.5}{8.5}\selectfont
\begin{eqnarray}
&rank(t_1)=\lfloor AM(5,2) \rfloor=3\nonumber\\
&rank(t_3)=\lfloor AM(3,4) \rfloor=3\nonumber\\
&rank(t_5)=\lfloor AM(1,3) \rfloor=2\nonumber\\
&rank(t_7)=\lfloor AM(4,1) \rfloor=2\nonumber\\
&rank(t_9)=\lfloor AM(2,5) \rfloor=3\nonumber\\
&\textnormal{\textit{AM-consensus}=$\langle[t_5,t_7],[t_1,t_3,t_9]\rangle$}\nonumber
\end{eqnarray}}

The initial approximation contains ties for ranks $1$ ($[t_5,t_7]$) and rank $2$ ($[t_1,t_3,t_9]$). Although in the context of this example the application of the floor ($\lfloor .\rfloor$) function makes the AM-consensus identical to the BC-consensus, practically they are different permutations even in the presence of ties.

\subsection{Test-suite execution (stage \mycircled{3} of 3)}\label{sec:execution}

\begin{figure}[h]
    \centering
    \includegraphics[width=0.5\linewidth]{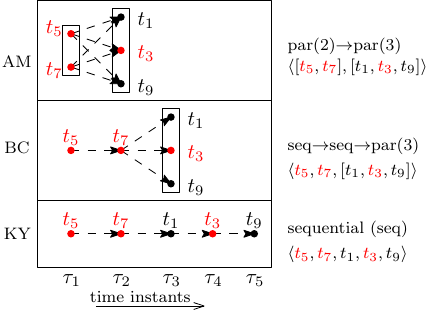}
    \caption{Execution of consensus prioritization as per aggregated preferences.}
    \label{fig:execution}
\end{figure}

The consensus prioritization of test cases may be executed sequentially or in a partially parallel manner depending on the presence of ties in the final consensus ordering. For our running example, the execution modes are illustrated in Figure~\ref{fig:execution} for AM, BC, and KY consensuses ($y$-axis). The execution timeline consists of five time instants: $\tau_{1..5}$ along the $x$-axis. The arrows represent preference relationship where $t_i\rightarrow{t_j}$ denotes $t_i$ executes-before $t_j$. Rectangular boxes containing test cases (circles) denote a parallelization window~\cite{mahtab_2019,hansie_2020,hansie_2021} which is a batch of multiple processes forked per test case. The ({\color{red}$\bullet$})/($\bullet$) circles represent failed/passed test cases, respectively when executed. 

\textit{Sequential execution}: As KY-consensus is strict ordering, test cases must be executed sequentially as per the consensus aggregated. Three test cases fail at positions 1, 2, and 3. Thus, with up to four test executions, all failures are observed (covered). However, note that the number of failures is never considered as a coverage criterion for prioritizing test cases in the first place. 

\textit{Parallel execution}: Although KY-consensus yields a total order, much looser orders exist via BC and AM consensuses. For BC, only the first two test cases need to be executed in sequence, and the rest three can be executed in parallel. The underlying assumption is that the SUT is re-entrant and all test cases are at the system level and hence, independent. This gives us a leeway to batch them in a batch of three parallel processes, hence address isolated and free from shared state conflicts (if any). The benefit thus obtained in doing so is that we cover all the failures at the end of $\tau_3$, i.e., third position in the execution timeline. Please note that the rank of the test cases ($t_1$, $t_3$, $t_9$) at $\tau_3$ is 3 (tied). Hence, if multiple test cases are tied for a particular rank, the corresponding sequential position in the execution timeline gets to observe that many test-executions in parallel (in principle), subject to availability of multiple cores (\textit{nproc}). For AM, \{$t_5$, $t_7$\} are executed in 2-parallel at time instant $\tau_1$. Next, \{$t_1$, $t_3$, $t_9$\} are executed in 3-parallel at time instant $\tau_2$. Hence, all the failures are covered within top two position of time execution timeline with AM-consensus. Should there be a parallelization window $w$ of size greater than \textit{nproc}, the entire $w$ should be chunked into sizes of \textit{nproc} (sub-windows). Next, the chunks (with intra-parallelism) are executed in sequence. The more the number of cores in the multi-core processor, the better is the conformation to the consensus ordering. This is turn also impacts the throughout in terms of the cumulative number of failures observed up to a time instant $\tau_i$ in the exectuion timeline.

\section{Experimental Setup}\label{sec:setup}

\subsection{Research Questions} \label{sec:questions}
The evaluation of \entp~\cite{replicationPackage} was driven by the following \textit{research questions}.
\begin{itemize}
	\item{\textbf{RQ1 (comparison against standalone)}: \textit{Do the consensus prioritization plans perform better than the standalone prioritizations?} (Section~\ref{sec:RQ1})}
	\item{\textbf{RQ2 (ensemble selection using diversity)}: \textit{Does the consensus of some prioritization plans perform better than the consensus of the entire ensemble?} (Section~\ref{sec:RQ2})}
	\item{\textbf{RQ3 (comparison against state-of-the-art)}: \textit{How beneficial is ensemble test prioritization compared to the state-of-the-art approaches?} (Section~\ref{sec:RQ3})}
\REM{	\item{\textbf{RQ4 (impact of paradoxes and ties)}: \textit{How does the presence of paradoxes and ties impact ensemble test prioritization?} (Section~\ref{sec:RQ4})}
 
	\item{\textbf{RQ5 (overhead of prioritization)}: \textit{What is the overhead of ensemble test prioritization?} The purpose of this RQ is to evaluate the overhead of the overall consensus mechanism (consensus computation + ensemble selection). If the overhead is significant, then even in the event of an effective consensus prioritization, the costs would out-weigh the benefits.} }
\end{itemize}

\subsection{Environment}
We performed our experiments on a computer powered by \Processor, \NumCPUs\ CPUs (\NumCores\ cores, with 2 threads per core), \RAMCapacity\ GB RAM, and with a storage of \HardDiskCapacity\ GB. Software used were \OSVersion\ with \CodeIn{Linux} kernel \KernelVersion, \CodeIn{gcc} \GCCVersion{}, \CodeIn{LLVM}  \LLVMVersion{}, and GNU \CodeIn{bash} \BashVersion.

\begin{table}[h]
	\centering
	\fontsize{8}{8}\selectfont
	\caption{Dataset attributes.}
	\label{tab:bench}
	\begin{tabular}{@{}l|rrrr|l@{}}
		\toprule
		{\textit{\textbf{Benchmark}}} & {\textit{\textbf{\#ver}}} & {\textit{\textbf{SLOC}}} & {\textit{\textbf{\#tests} (system level)}} & {\textit{\textbf{\#faults, fault-type}}} & {\textit{\textbf{URL (https://)}}} \\ \midrule
		\texttt{c4} & 11  & 4640 & 1170 & unknown, real & \url{github.com/rswier/c4} \\
		\texttt{ckf} & 11 & 4994 & 222 & unknown, real & \url{github.com/.../CuckooFilter} \\
		\texttt{flex} & 5 &  92100 & 670 & 62 (0+20+17+16+9), real & \url{sir.csc.ncsu.edu/content} \\
		\texttt{gravity} & 11 & 45612 & 923 & unknown, real & \url{github.com/marco.../gravity} \\
		\texttt{grep} & 6 & 102084 & 809 & 57 (0+18+8+18+12+1), real & \url{sir.csc.ncsu.edu/content} \\
		\texttt{gzip} & 6 & 54043 & 214 & 59 (0+16+7+10+12+14), real & \url{sir.csc.ncsu.edu/content} \\
		\texttt{mlisp} & 6 & 4238 & 3721 & unknown, real & \url{github.com/rui314/minilisp} \\
		\texttt{printtokens} & 8 & 2433 & 4130 & 7 (0+1*7), seeded & \url{sir.csc.ncsu.edu/content} \\
		\texttt{printtokens2} & 11 & 3945 & 4115 & 10 (0+1*10), seeded & \url{sir.csc.ncsu.edu/content} \\
		\texttt{replace} & 33 & 16893 & 5542 & 32 (0+1*32), seeded & \url{sir.csc.ncsu.edu/content} \\
		\texttt{scd} & 11 & 15126 & 9261 & unknown, real & \url{github.com/.../sha1collision...} \\
		\texttt{schedule} & 10 & 2918 & 2650 & 9 (0+1*9), seeded & \url{sir.csc.ncsu.edu/content} \\
		\texttt{schedule2} & 11 & 2913 & 2710 & 10 (0+1*10), seeded & \url{sir.csc.ncsu.edu/content} \\
		\texttt{sed} & 5 & 63241 & 370 & 18 (0+3+5+6+4), real & \url{sir.csc.ncsu.edu/content} \\
		\texttt{slre} & 9 & 3247 & 132 & unknown, real & \url{github.com/cesanta/slre} \\
		\texttt{space} & 39 & 230723 & 13585 & 38 (0+1*38), real & \url{sir.csc.ncsu.edu/content} \\
		\texttt{tcas} & 42 & 5672 & 1608 & 41 (0+1*41), seeded & \url{sir.csc.ncsu.edu/content} \\
		\texttt{totinfo} & 24 & 6676 & 1052 & 23 (0+1*23). seeded & \url{sir.csc.ncsu.edu/content} \\
		\texttt{xc} & 11 & 10476 & 800 & unknown, real & \url{github.com/.../...C-interpreter} \\
		\texttt{xxhash} & 10 & 22538 & 15621 & unknown, real & \url{github.com/Cyan.../xxHash} \\\midrule
		{\textit{\textbf{Total}}} & \textit{\textbf{280}} & \textit{\textbf{694512}} & \textit{\textbf{69305}} & \textit{\textbf{---}} & \textit{\textbf{---}}  \\ \bottomrule
	\end{tabular}
\end{table}

\begin{figure*}[h]
    \centering
    \fbox{\includegraphics[width=0.98\linewidth]{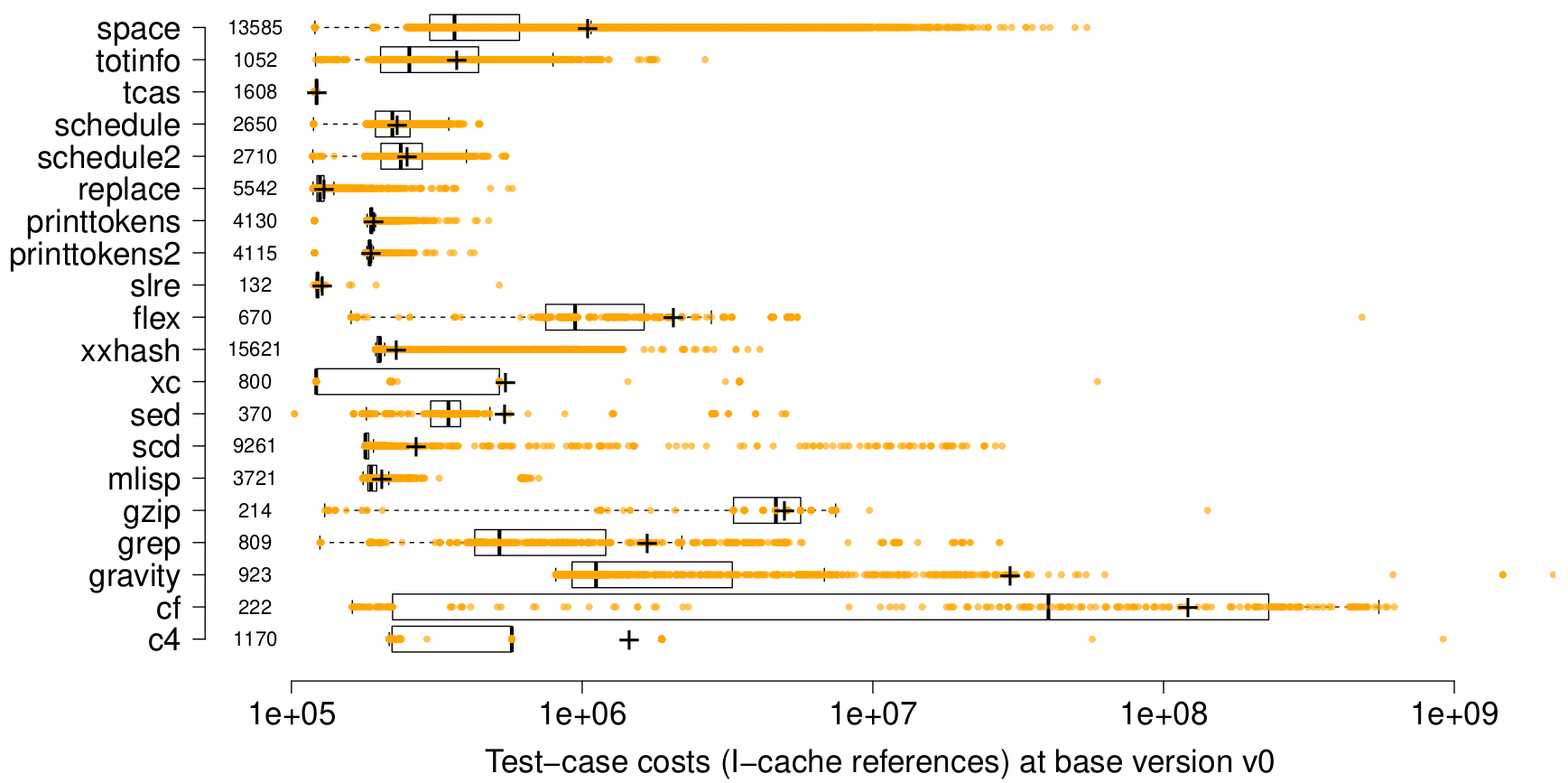}}
    \caption{Distribution of test case costs within the full test-suite at base version $v_0$ for our dataset. (\textit{boxplot}: a datapoint represents a test case).}
    \label{fig:load_distribution_v0}
\end{figure*}

\begin{figure}[h]
    \centering
    \fbox{\includegraphics[width=0.78\linewidth]{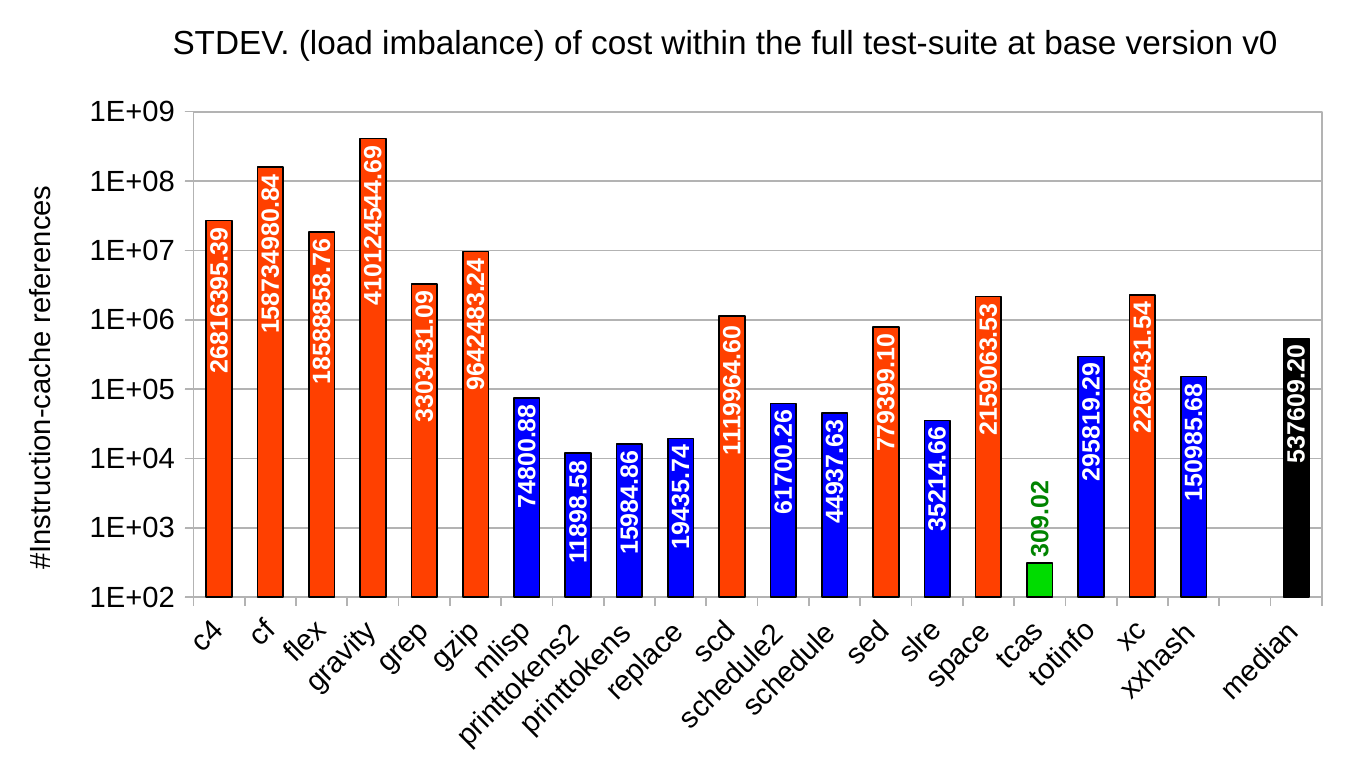}}
    \caption{Test case load (cost) imbalance within respective test-suites for the studied objects.}
    \label{fig:load_imbalance}
\end{figure}

\REM{
\begin{figure}[h]
    \centering
    \fbox{\includegraphics[width=0.98\linewidth]{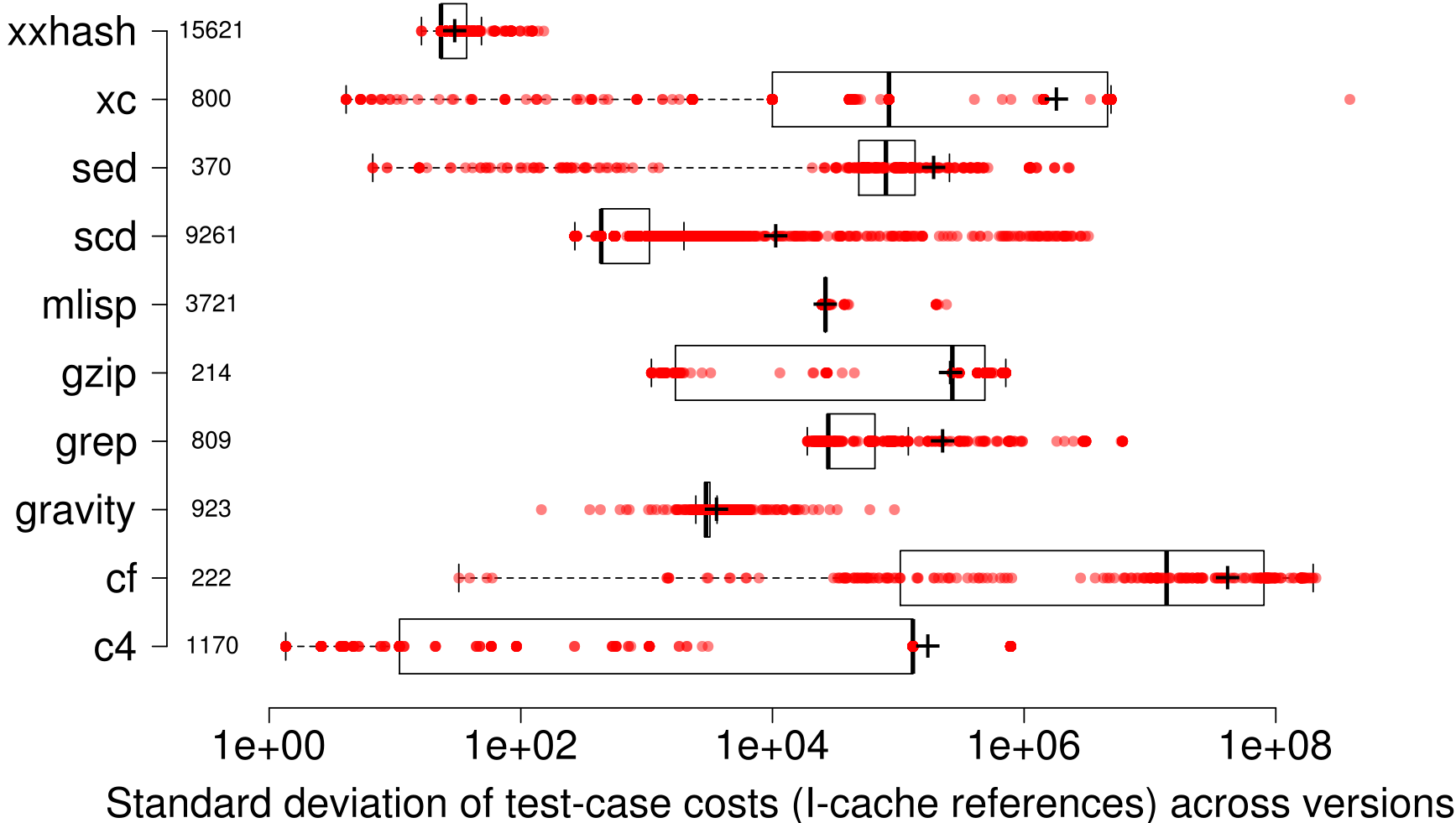}}
    \caption{Distribution of test case load imbalance (s.d. of cost) across versions for the studied objects. (\textit{boxplot}: a datapoint represents a test case).}
    \label{fig:load_distribution_all}
\end{figure}
}
\subsection{Benchmarks}
\entp was evaluated using twelve programs from the Software-artifact Infrastructure Repository (SIR)~\cite{do_2005}, and eight projects from GitHub. Attributes of these benchmarks are shown in Table~\ref{tab:bench}, columns are sorted lexicographically by benchmark name. The table is described as follows.

\subsubsection{Selection criteria}
Twelve subjects were selected from the popular SIR repository, and eight projects were picked from GitHub. An objective selection criteria~\cite{mahtab_2019,hansie_2020,col_tse_2021} was used: (i) language distribution: C ($100\%$), (ii) \texttt{gcc}-based command-line compilation, (iii) project does not rely on build systems such as \texttt{make}, \texttt{cmake}, etc. as these are yet to be supported in our implementation. All subjects and test cases with known flaky behavior were excluded~\cite{Cheng_2021} as those may pass or fail even without any code or test changes, in which case the valuation of effectiveness measures will be non-deterministic and unreliable for evaluation. 
Finally, we settled on 20 benchmarks presently supported by our current implementation of \entp.

\subsubsection{Versions per benchmark}
Total number of versions (\textit{\textbf{\#ver}}) for each benchmark are shown in the second column. Note that \hansie~\cite{hansie_2020} used the same set of benchmarks. However, the authors evaluated with only 4-5 versions per subject. In contrast to this, \entp has been evaluated on the dataset in its entirety, for all available versions. Thus, \entp competes with \hansie on a broader evaluation aspect.

\subsubsection{Code change (affected basic-blocks)}
We transformed each project into a single LLVM IR (\texttt{.ll} file) by passing the \texttt{-O0 -S -emit-llvm} flags to \texttt{clang}. Next, the Unix \texttt{diff} on two versions of LLVM IRs. A basic-block was considered affected jointly with respect to both versions if it was newly introduced/added (in the new version), modified, or deleted/removed (from the older version). The \textit{def-use} of global variables were also tracked from the LLVM IRs and was treated as affected as per the aforementioned criteria for basic-blocks. Therefore, we also considered a basic block affected if it \textit{use}d any affected (added/modified/deleted) global variable. We followed the granularity of basic-blocks (code elements) throughout our implementations in \entp.

\subsubsection{Scale of benchmarks, test-sets, and faults}
For each benchmark, the total count of physical source lines of code (\textit{\textbf{SLOC}}) across all versions, size of the test-pool (system level \textbf{\textit{\#tests}}), and the fault multiplicity/type (\textit{\textbf{\#faults, fault-type}}) are shown in their respective columns. URL of each benchmark in their original (unprocessed) form appears in the right-most column (\textbf{\textit{URL}}). The bottom row displays the sum total of respective attributes for our studied dataset. 

\subsection{Oracles and regression test case verdicts}
In this work, we considered only system level test cases. A regression test case was considered to have failed if the expected output corresponding to the clean version matched exactly with the observed output in the SUT. Our interest was bound to \textit{detection of failures}. Thus, in \entp, the prioritization was heuristically guided to improve the \textit{rate of failure observation}. This approximation of faults by failures had been advocated by prior research~\cite{FAST_2018,Chen_2018,Yu_2019}. We adopt this simplifying approximation for two reasons. Firstly, the fault knowledge in not available before the test executions. Secondly, a failure event may possibly arise in the presence of multiple faults in which case it is difficult to know what one actually led to the error (failed) state.

\subsection{Distribution of test case costs within test-suites}\label{sec:imbalance}
As raw execution time is noisy, we approximated the cost of a test case (executed on the old version) by the number of I-cache references made (event \texttt{Ir}) as profiled by \texttt{cachegrind}~\cite{cachegrind} while running the test case. We assumed static test-suite which do not evolve across versions. Hence, the distribution of the cost at the base version $v_0$ is of prime importance for test case prioritization as well as parallelization. Figure~\ref{fig:load_distribution_v0} shows this distribution and reflects that test costs vary significantly (as shown in Figure~\ref{fig:load_imbalance}) within test suites for 19/20 (95\%) of the benchmarks, the exception being \texttt{tcas} with (mean: 121956.02, median: 122014, s.d.: 309.02) which is relatively low. The reason for this is the simplistic nature of test cases (total 1608) which are a few command line integers and do not involve disk I/O. The only other benchmark with user input via \texttt{scanf} only is \texttt{slre} which accepts a number (as a test case) from 1 to 132. However, it has a \texttt{switch-case} construct for different operations. Five out of 132 test cases in \texttt{slre} have relatively unbalanced execution costs (131693, 158054, 195518, 518762, 161285), all others have costs ranging from 12200 to 12500. These give \texttt{slre} a test cost distribution with (mean: 127342.83, median: 122834, s.d.: 35214.66). As extremes: (i) \texttt{gravity} has the largest load imbalance with 923 test cases, and (ii) \texttt{cf} has the highest median test cost 40169817.5 within its test suite of 222 test cases consuming a disk space of 329.77 MB (1.49 MB per test case). Overall, our dataset has three classes of benchmarks with high, medium, and low standard deviation of test case costs within their respective test suites. The corresponding bar plots are shaded in {\color{red}\textbf{red}}, {\color{blue}\textbf{blue}}, and {\color{indiagreen}\textbf{green}} (Figure~\ref{fig:load_imbalance}). The median load imbalance (across benchmarks) in the studied dataset was 537609.20. 

\REM{
\subsection{Compared non-consensus based prioritization strategies}\label{sec:baselines} 

We compared \entp against three state-of-the-art prioritizations: (i) \textit{greedy total (relevance)}~\cite{APFD_2001,APFD_2002,mahtab_2019}, (ii) \textit{greedy additional}~\cite{APFD_2001,APFD_2002}, and (iii) ...  Additionally, we compared against a recently proposed coverage-based prioritization: prioritization by \textit{relevance-and-confinedness}~\cite{mahtab_2019}.

\subsubsection*{Relevance-and-confinedness based approach} This is a recently proposed delta-coverage based prioritization scheme that heuristically considers a test case to have a high priority if it has \textit{high delta and low non-delta} coverage. Henceforth, we refer to this prioritization as \textit{relcon}.
}

\subsection{Measuring the effectiveness of test case prioritization}\label{sec:effectiveness}
We provide a concise overview of the metrics employed to incentivize efficient prioritization of test case executions. The metric values ranges between 0 (bad) and 1 (good), both inclusive.
 
\paragraph*{\textbf{APFD}}
This metric quantifies the \textit{average percentage of failures detected}~\cite{APFD_2001,APFD_2002} throughout the \textit{timeline} of test execution. Geometrically, it corresponds to the area under the curve (AUC) formed by the cumulative percentage of failures observed on the $y$-axis and the percentage of the executed test suite on the $x$-axis.

\paragraph*{\textbf{APFD$_{\textit{\textbf{c}}}$}}
This metric employed in this study is a \textit{cost-aware adaptation of APFD}\cite{APFD_c}, which incorporates the execution time (cost) of regression test cases in the version-under-test. Furthermore, this metric also takes into account non-uniform fault severity. We adopt the assumption from Epitropakis et al.\cite{Epitropakis_2015} that all test failures are of equal severity.


\paragraph*{\textbf{EPS}}
This metric proposed by Mondal and Nasre~\cite{mahtab_2019} has a distinct foundation compared to the previously mentioned metrics. Its effectiveness is determined by the normalized Manhattan distance between the observed binary failure vector (with ``0" denoting passed and ``1" denoting failed) resulting from prioritized test execution, and a hypothetical optimal pattern where an equal number of test failures occur consecutively, starting from the first position along the execution timeline. The similarity, denoted as \textit{EPSilon}, is calculated by taking the complement of the computed distance within the interval of $[0,1]$.

\section{Experimental Results}\label{sec:exp}
In this section, we address the research questions formulated in Section~\ref{sec:questions} through our evaluation.

\subsection{Answering RQ1 (comparison against standalone)}\label{sec:RQ1}

\noindent
\textbf{Motivation.} In this RQ, we evaluate whether consensus prioritization by rank aggregation indeed brings benefit in terms of failure throughput. If standalone heuristics always turn out to be more effective, then consensus should be avoided.

\noindent
\textbf{Approach.} We considered 16 standalone prioritizations (as in Table~\ref{tab:strategies}) participating in the ensemble, and six different consensus approaches: Borda-count (borda), Kemeny-Young (ky), Arithmetic-Mean (am), Geometric-Mean (gm), Harmonic-Mean (hm), Median (med), along with four choice of top-$k\%$ budgets: 100\%, 75\%, 50\%, 25\%. Across these configurations, the number of distinct prioritizations turned out to be 40 (16+6*4). Consensus strategies labeled as ``\{\texttt{borda/ky/am/gm/hm/med\}\_100}" denote \hansie (as well as \entp by extension), otherwise labels (for consensus strategies) suffixed with \texttt{\_\{25,50,75\}} denote \entp exclusively.

\begin{figure}[h]
    \centering
    \includegraphics[width=\linewidth]{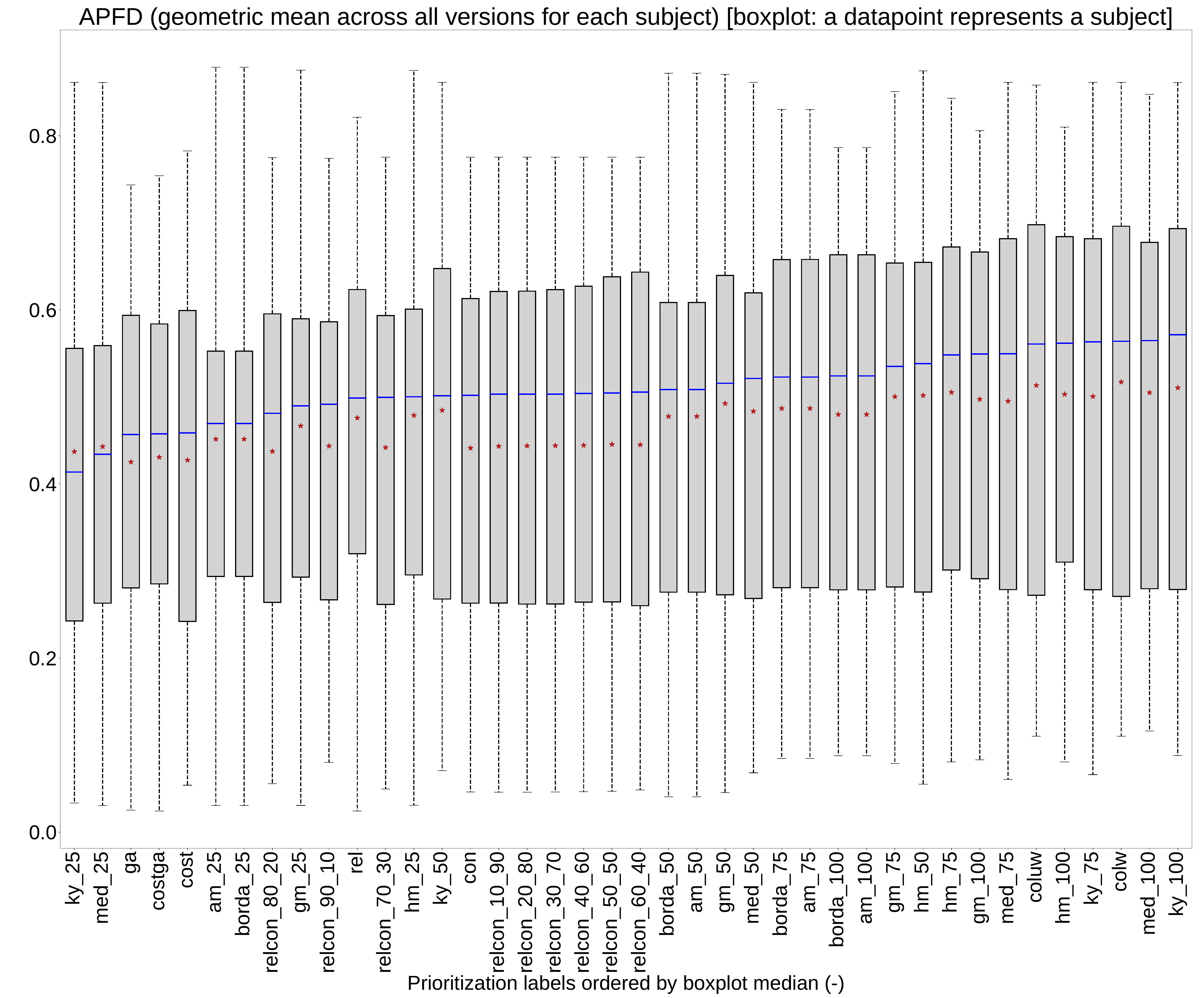}
    \caption{Effectiveness of prioritizations: APFD values.}
    \label{fig:metrics_APFD}
\end{figure}

\begin{figure}[h]
    \centering
    \includegraphics[width=\linewidth]{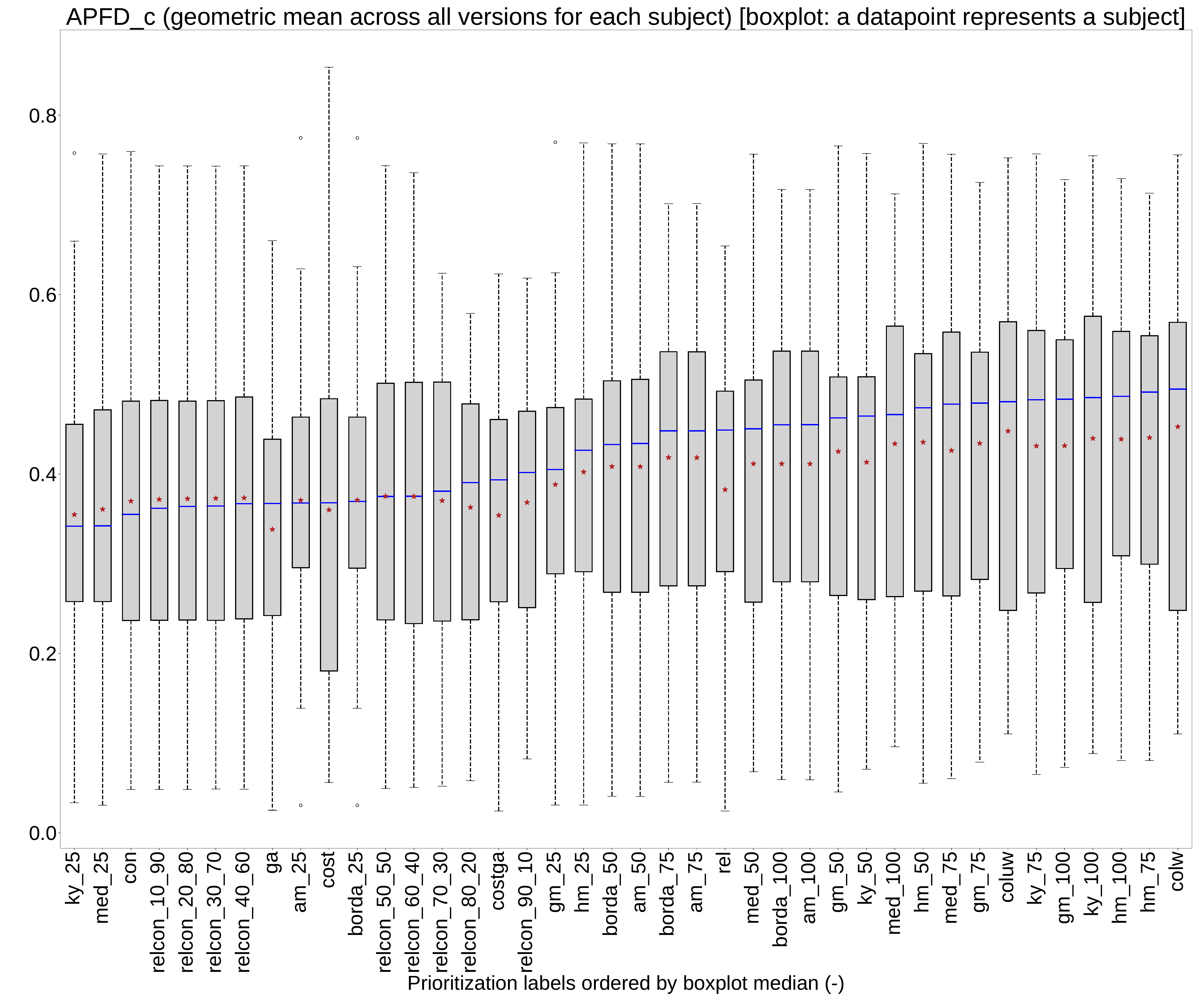}
    \caption{Effectiveness of prioritizations: APFD$_c$ values.}
    \label{fig:metrics_APFDc}
\end{figure}

\begin{figure}[h]
    \centering
    \includegraphics[width=\linewidth]{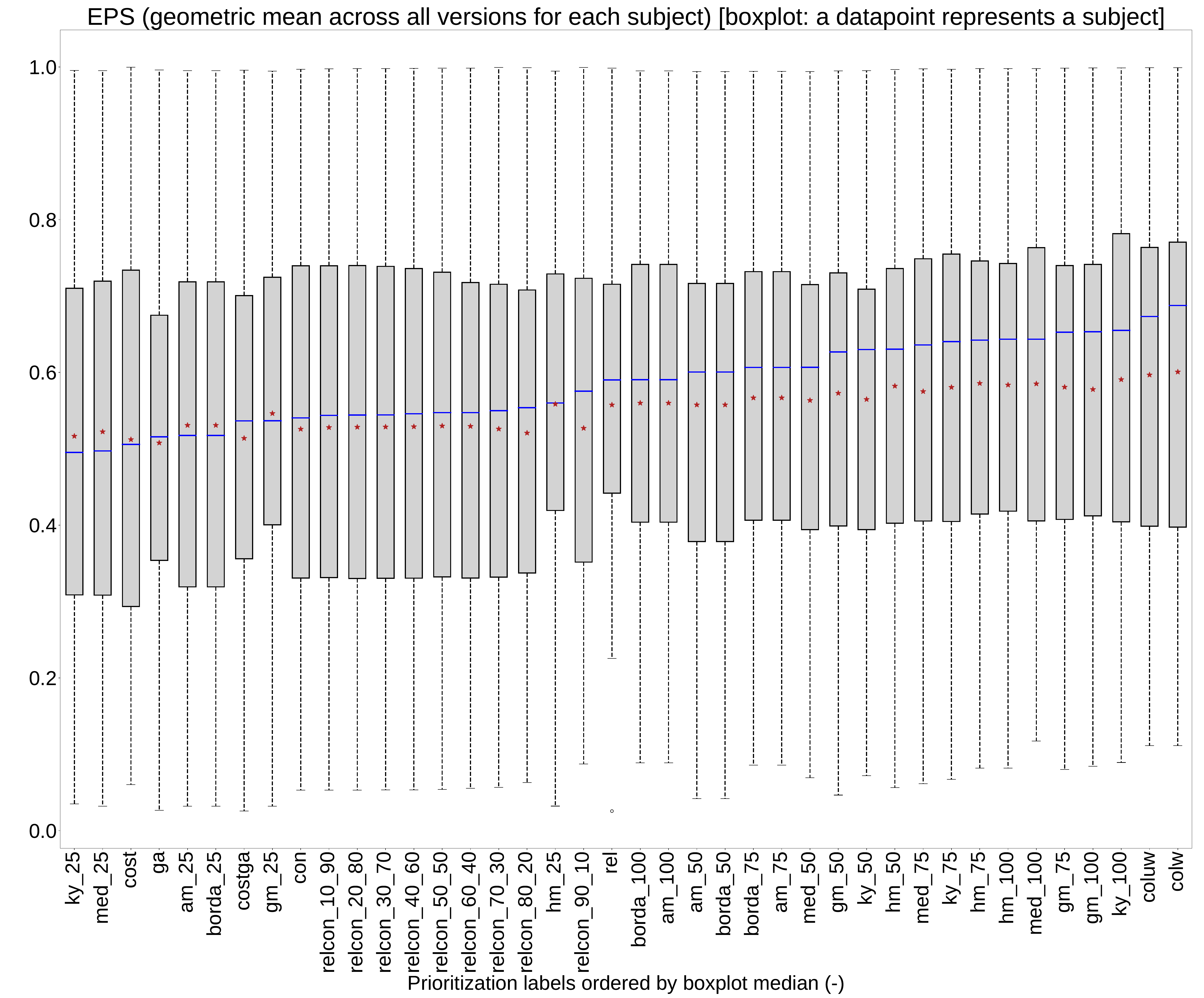}
    \caption{Effectiveness of prioritizations: EPS values.}
    \label{fig:metrics_EPS}
\end{figure}

\noindent {\bf Results.} Figures~(\ref{fig:metrics_APFD},~\ref{fig:metrics_APFDc},~\ref{fig:metrics_EPS}) show the boxplots of our chosen effectiveness metrics: APFD, APFD$_c$, and EPS. For these plots, the $x$-axes represent prioritization labels ordered by better effectiveness (left to right), and the $y$-axes denote the corresponding metric values, the closer to 1 the better. We observe that the consensus strategies (both \hansie and \entp) with at least one configuration outperform standalone prioritizations for APFD. For APFD$_c$ and EPS, \colosseum (\texttt{colw}) outperformed all single-handedly. The overall performance of all consensus strategies (total 24) spanning Figures~(\ref{fig:metrics_APFD},~\ref{fig:metrics_APFDc},~\ref{fig:metrics_EPS}) is summarized in Table~\ref{tab:standalone}. The left column for each metric shows the rank of a strategy among the 40 combinations (standalone as well as consensus), while the right column shows how many of the standalone prioritizations (total 16) were outperformed (i.e. defeated) by a consensus strategy. For instance, in the space of 40 prioritizations, \texttt{ky\_75} (Kemeny-Young consensus of top 75\% diverse prioritizations), was the fourth, sixth, and ninth best in terms of APFD, APFD$_c$, and EPS, respectively. Compared against 16 standalones, \texttt{ky\_75} was able to outperform 15/16 of them, all but \texttt{colw}. Similar explanation holds for other columns as well. 

\begin{table}[h]
\caption{Rank of different consensus strategies and their defeat strength against standalone prioritizations.}
\centering
\fontsize{8}{8}\selectfont
\begin{tabular}{@{}l|r|r|r|r|r|r@{}}
\toprule
\multirow{2}{*}{\textbf{\textit{Strategy}}} &
  \multicolumn{2}{c|}{\textbf{\textit{APFD}}} &
  \multicolumn{2}{c|}{\textbf{\textit{APFD$\bm{_c}$}}} &
  \multicolumn{2}{c}{\textbf{\textit{EPS}}} \\
  &
  \multicolumn{1}{l}{\textbf{\textit{Rank (1--40)}}} &
  \multicolumn{1}{l|}{\textbf{\textit{Defeat (\%)}}} &
  \multicolumn{1}{l}{\textbf{\textit{Rank (1--40)}}} &
  \multicolumn{1}{l|}{\textbf{\textit{Defeat (\%)}}} &
  \multicolumn{1}{l}{\textbf{\textit{Rank (1--40)}}} &
  \multicolumn{1}{l}{\textbf{\textit{Defeat (\%)}}} \\
  \midrule
\textbf{borda\_100} & 13 & 87.5  & 15 & 87.5  & 20 & 87.5  \\
\textbf{borda\_75}  & 15 & 87.5  & 19 & 81.25 & 16 & 87.5  \\
\textbf{borda\_50}  & 19 & 87.5  & 21 & 81.25 & 17 & 87.5  \\
\textbf{borda\_25}  & 34 & 18.75 & 30 & 43.75 & 35 & 12.5  \\
\textbf{ky\_100}    & 1  & 100   & 4  & 93.75 & 3  & 87.5  \\
\textbf{ky\_75}     & 4  & 93.75 & 6  & 93.75 & 9  & 87.5  \\
\textbf{ky\_50}     & 27 & 43.75 & 12 & 87.5  & 12 & 87.5  \\
\textbf{ky\_25}     & 40 & 0     & 40 & 0     & 40 & 0     \\
\textbf{am\_100}    & 12 & 87.5  & 14 & 87.5  & 19 & 87.5  \\
\textbf{am\_75}     & 14 & 87.5  & 18 & 81.25 & 15 & 87.5  \\
\textbf{am\_50}     & 18 & 87.5  & 20 & 81.25 & 18 & 87.5  \\
\textbf{am\_25}     & 35 & 18.75 & 32 & 37.5  & 36 & 12.5  \\
\textbf{gm\_100}    & 8  & 87.5  & 5  & 93.75 & 4  & 87.5  \\
\textbf{gm\_75}     & 11 & 87.5  & 8  & 87.5  & 5  & 87.5  \\
\textbf{gm\_50}     & 17 & 87.5  & 13 & 87.5  & 13 & 87.5  \\
\textbf{gm\_25}     & 32 & 25    & 23 & 81.25 & 33 & 18.75 \\
\textbf{hm\_100}    & 5  & 93.75 & 3  & 93.75 & 7  & 87.5  \\
\textbf{hm\_75}     & 9  & 87.5  & 2  & 93.75 & 8  & 87.5  \\
\textbf{hm\_50}     & 10 & 87.5  & 10 & 87.5  & 11 & 87.5  \\
\textbf{hm\_25}     & 28 & 43.75 & 22 & 81.25 & 23 & 75    \\
\textbf{med\_100}   & 2  & 100   & 11 & 87.5  & 6  & 87.5  \\
\textbf{med\_75}    & 7  & 87.5  & 9  & 87.5  & 10 & 87.5  \\
\textbf{med\_50}    & 16 & 87.5  & 16 & 87.5  & 14 & 87.5  \\
\textbf{med\_25}    & 39 & 0     & 39 & 0     & 39 & 0     \\
\bottomrule

\end{tabular}%
\label{tab:standalone}
\end{table}

{\fontsize{9}{9}\selectfont
	\begin{tcolorbox}
		{\textbf{\textit{\underline{Answering RQ1 (comparison against standalone)}}}}
		\begin{itemize}
			\item{\entp turns out to be a good consensus prioritization strategy in terms of cost cognizance effectiveness (APFD$_c$) with test suites having high load imbalance (overall median: 537609.20 I-cache references).}
                \item{\entp configured with a diversity budget of top-75\% for Kemeny-Young (KY), Harmonic Mean (HM), Geometric Mean (GM) consensus outperforms 93.75\%, 93.75\%, 87.5\% standalone prioritizations in isolation in terms of APFD, APFD$_c$, and EPS metrics, respectively.}
		\end{itemize}
	\end{tcolorbox}
}

\subsection{Answering RQ2 (ensemble selection using diversity)}\label{sec:RQ2}

\noindent
\textbf{Motivation.} In this RQ, we evaluate whether taking the consensus of some diverse prioritization may yield more benefit than a blind consensus (as performed by state-of-the-art \hansie). Consensus is more appropriate when there is significant diversity~\cite{compsoc}.

\begin{figure}[h]
    \centering
    \includegraphics[width=0.44\linewidth]{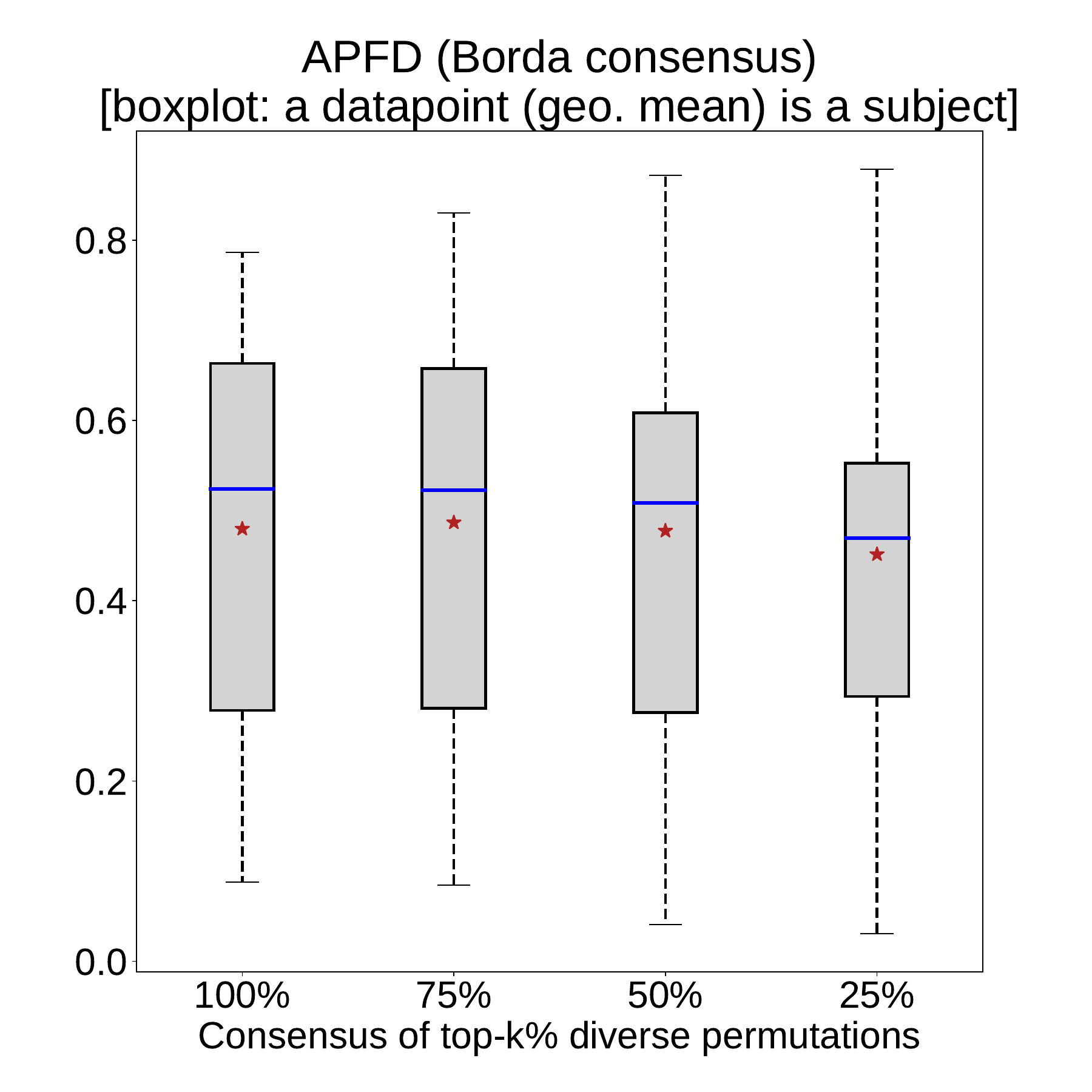}
    \includegraphics[width=0.44\linewidth]{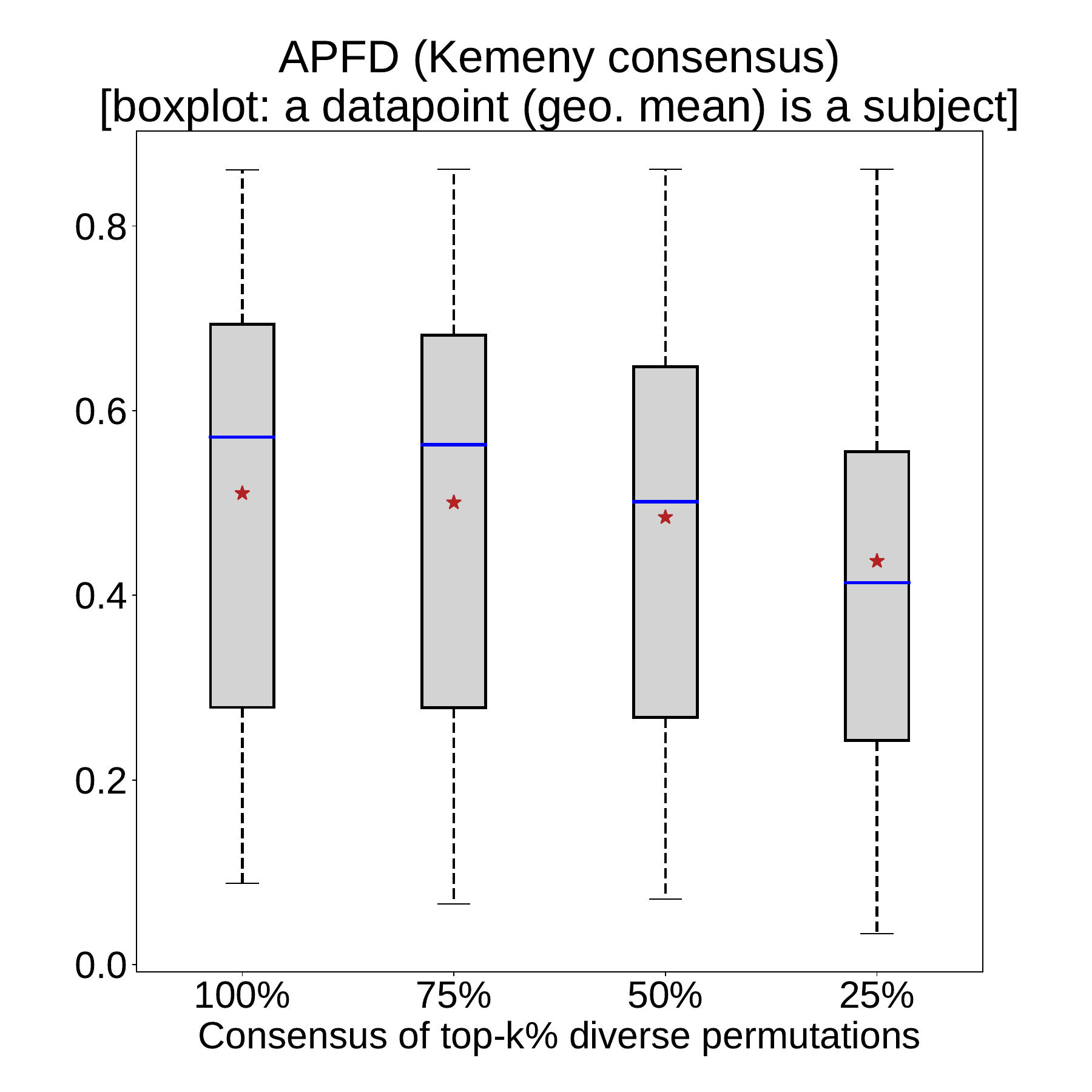}
    \caption{Ensemble selection using diversity: APFD values.}
    \label{fig:diversity_APFD}
\end{figure}

\begin{figure}[h]
    \centering
    \includegraphics[width=0.44\linewidth]{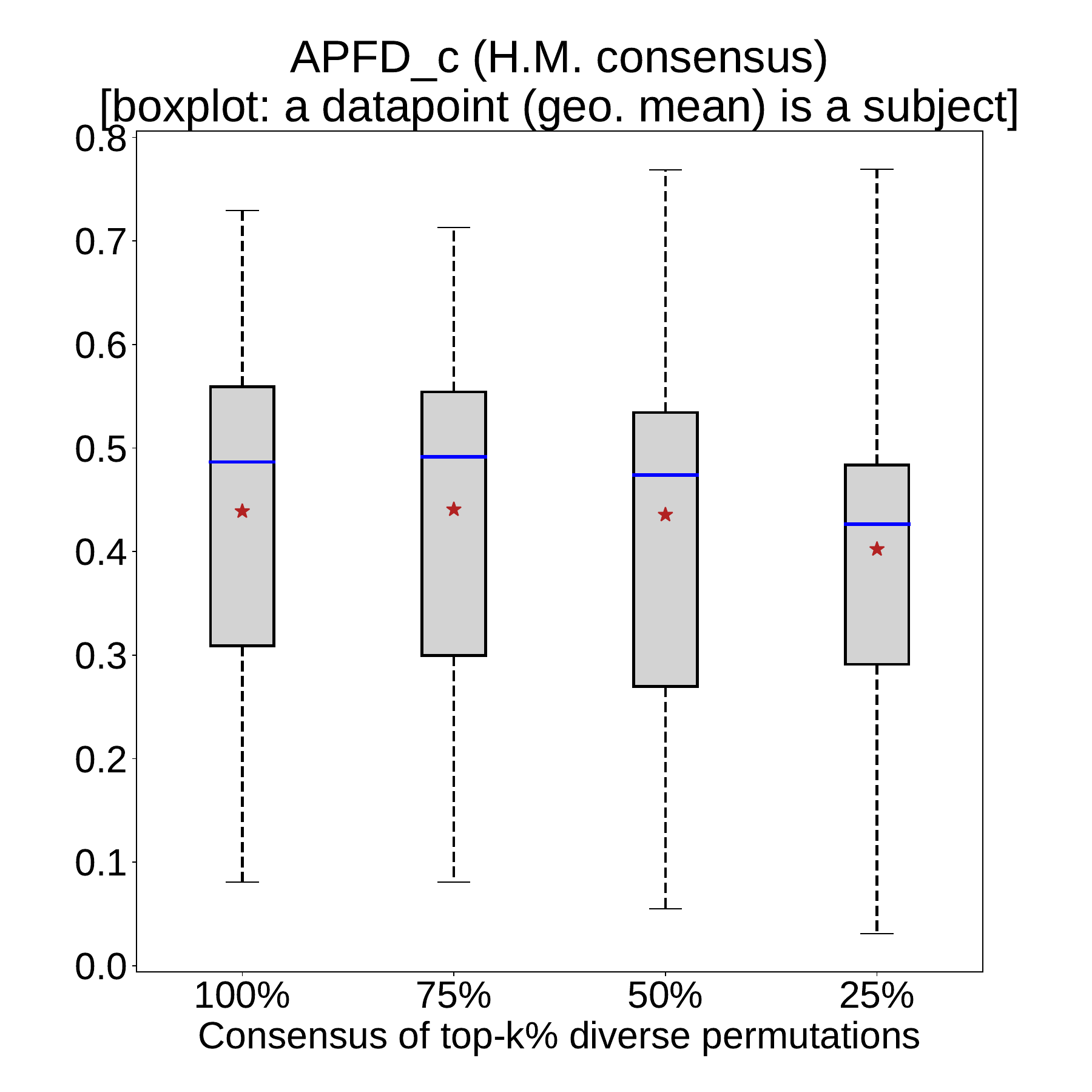}
    \includegraphics[width=0.44\linewidth]{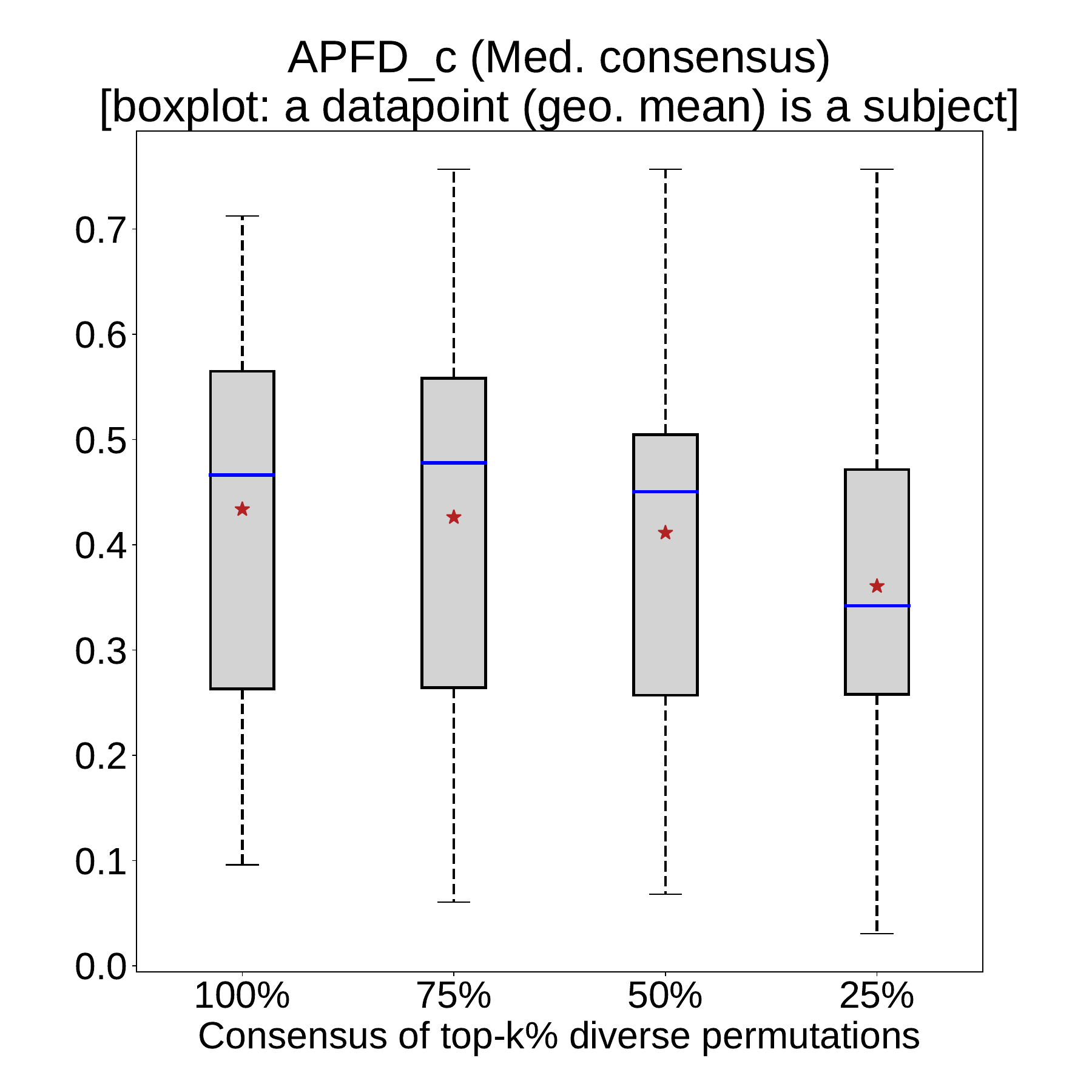}
    \caption{Ensemble selection using diversity: APFD$_c$ values.}
    \label{fig:diversity_APFDc}
\end{figure}

\begin{figure}[h]
    \centering
    \includegraphics[width=0.44\linewidth]{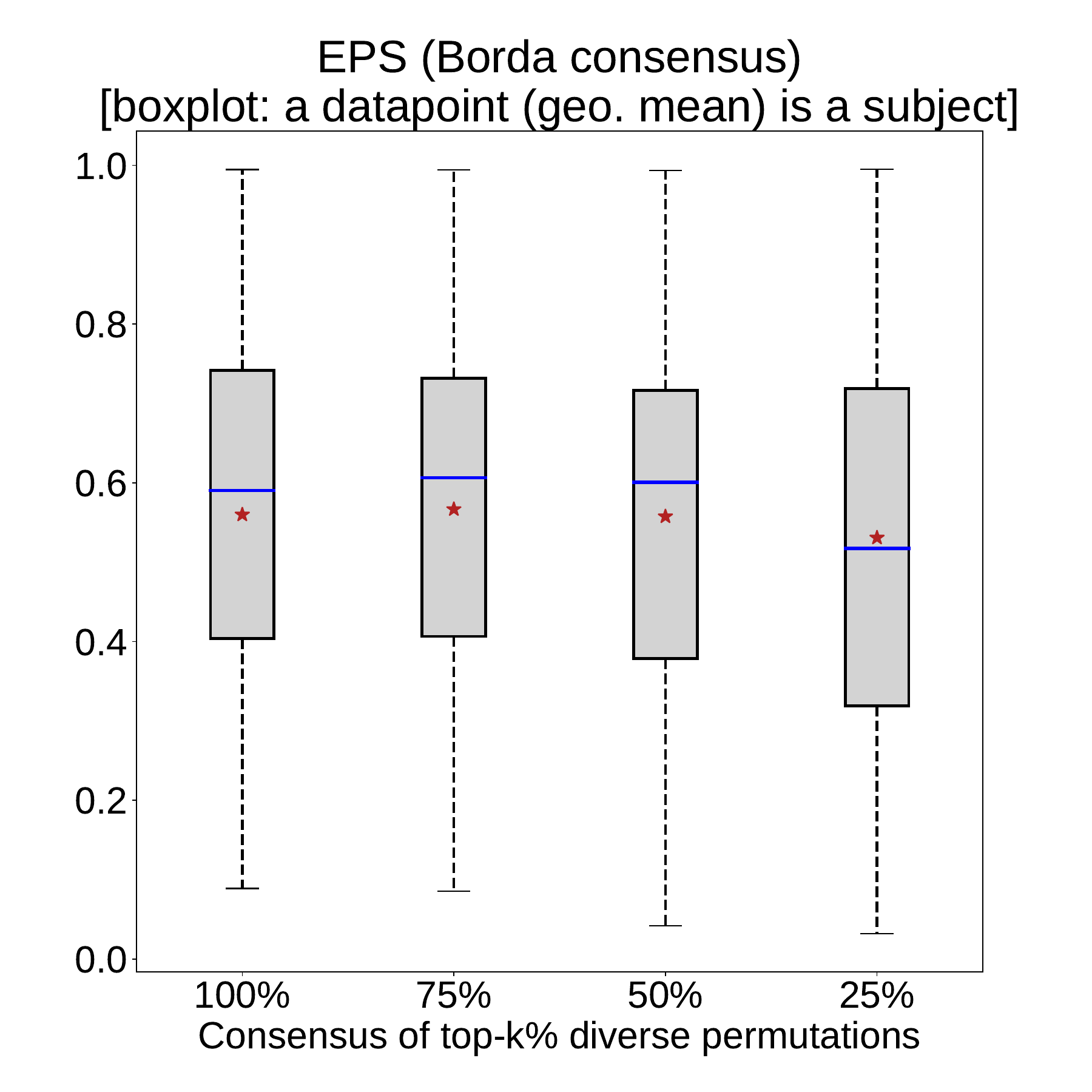}
    \includegraphics[width=0.44\linewidth]{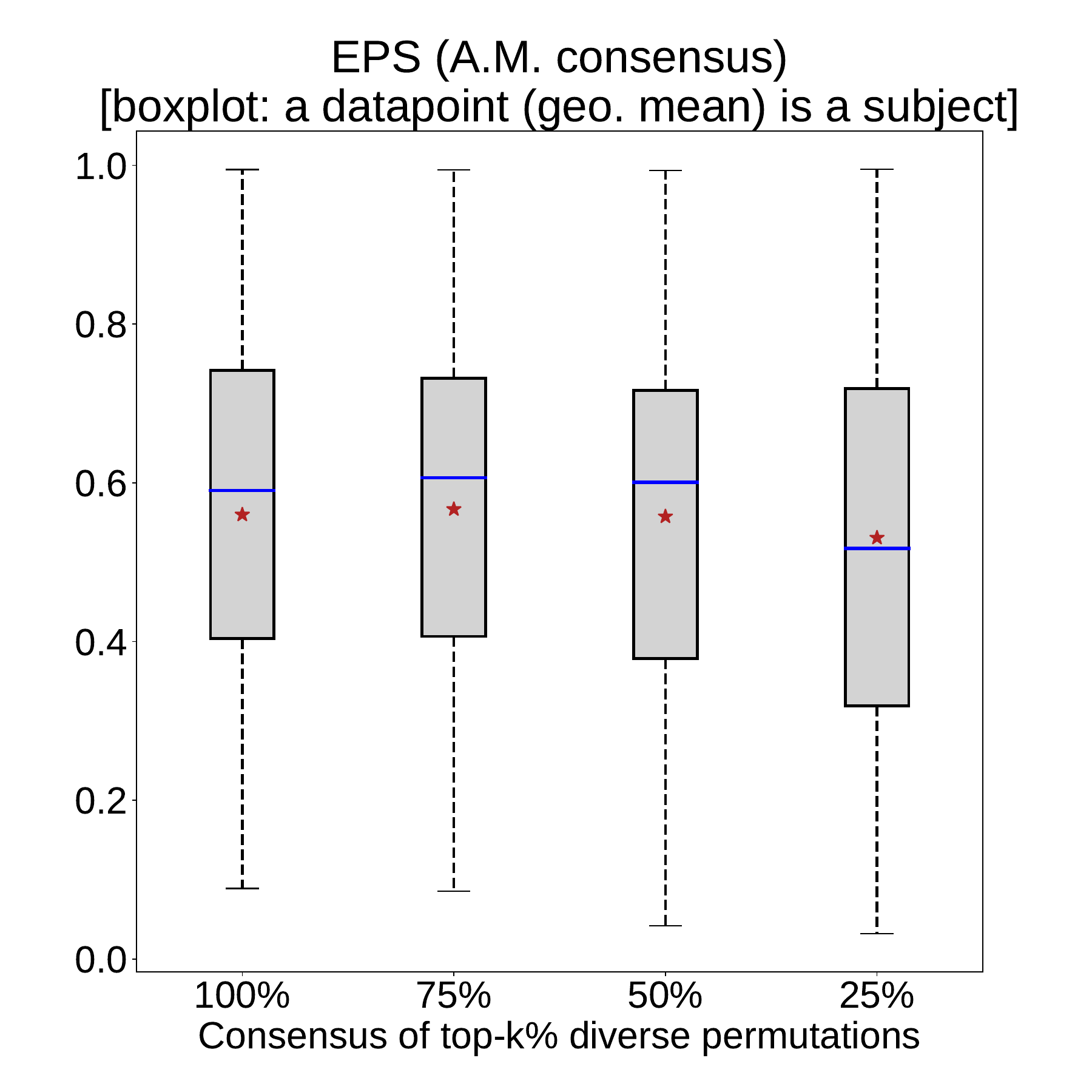}
    \caption{Ensemble selection using diversity: EPS values.}

    \label{fig:diversity_EPS}
\end{figure}

\noindent
\textbf{Approach and Results.} To gauge the benefit of diversity based ensemble selection, followed by top-$k\%$ budgeted rank aggregation, we perform comparison \hansie (k=100\%) and \entp (k=75\%, 50\%, 25\%). This is shown in the boxplots in Figures~\ref{fig:diversity_APFD},~\ref{fig:diversity_APFDc},~\ref{fig:diversity_EPS} for the associated effectiveness metrics. We observe that although in general, the effectiveness decreases with a decrease in budget, the reason for this can be attributed (in terms of APFD) to a few closely spaced good prioritizations otherwise lost due to diversity. However, if effectiveness is measured with test costs considered non-identical as is the case in the real world and in our dataset, the diversity budget of top-75\% is good enough to outperform a full blind consensus (\hansie). We measured the joint time taken: (i) to perform the diversity computation, (ii) ensemble selection. We found no significant difference (median range: 0.31--0.33 secs.) among the non-KY configurations (irrespective of budgets). However, \texttt{ky\_25} (median: 0.86 secs.), \texttt{ky\_\{50,75,100\}} (median: 0.81, 0.81, 0.87 secs.) show that Kemeny-Young consensus (irrespective of diversity budgets) turns out to be inefficient because of higher time complexity.

\begin{figure*}[h]
    \centering
    \includegraphics[width=0.495\linewidth]{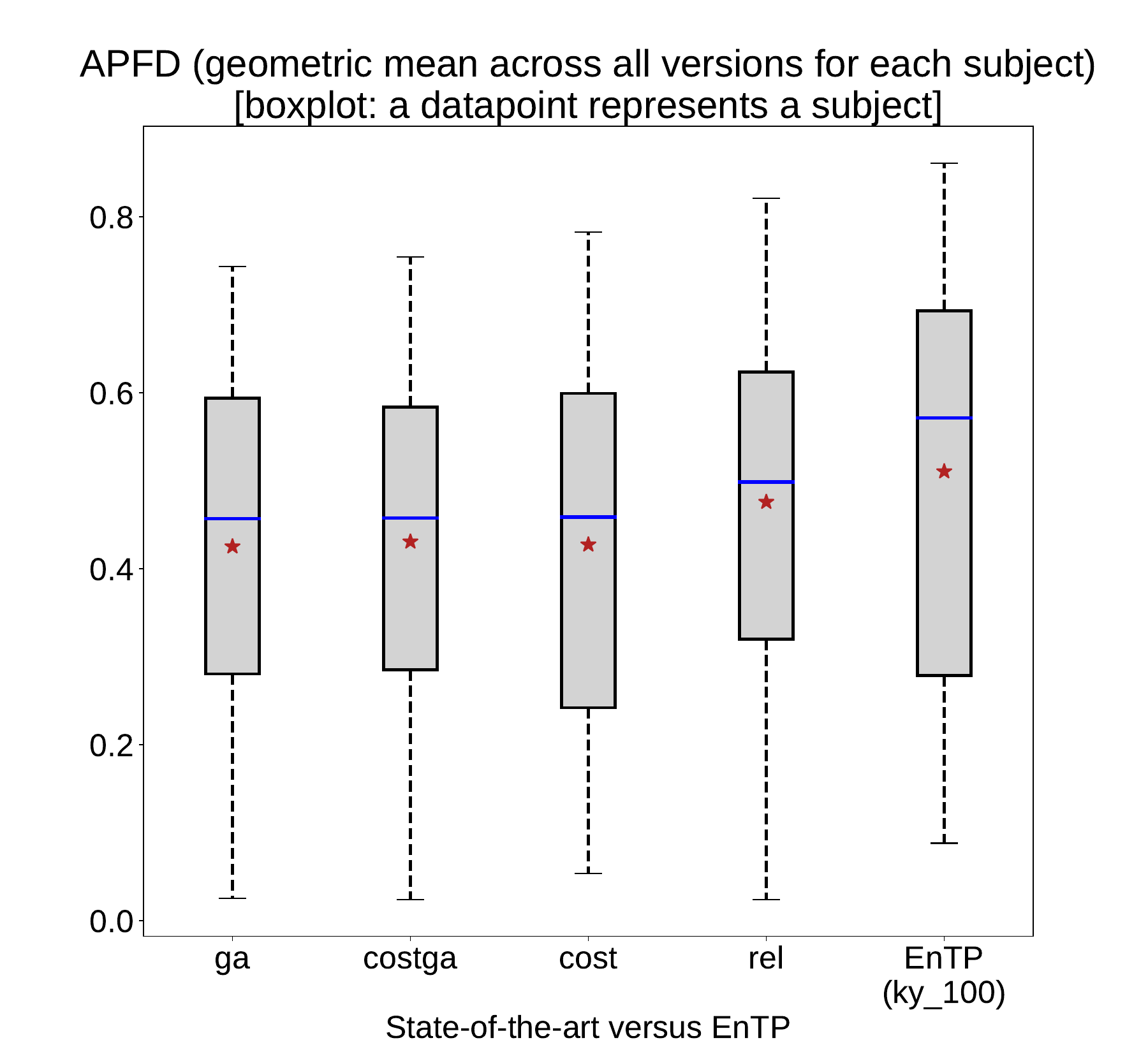}
    \includegraphics[width=0.495\linewidth]{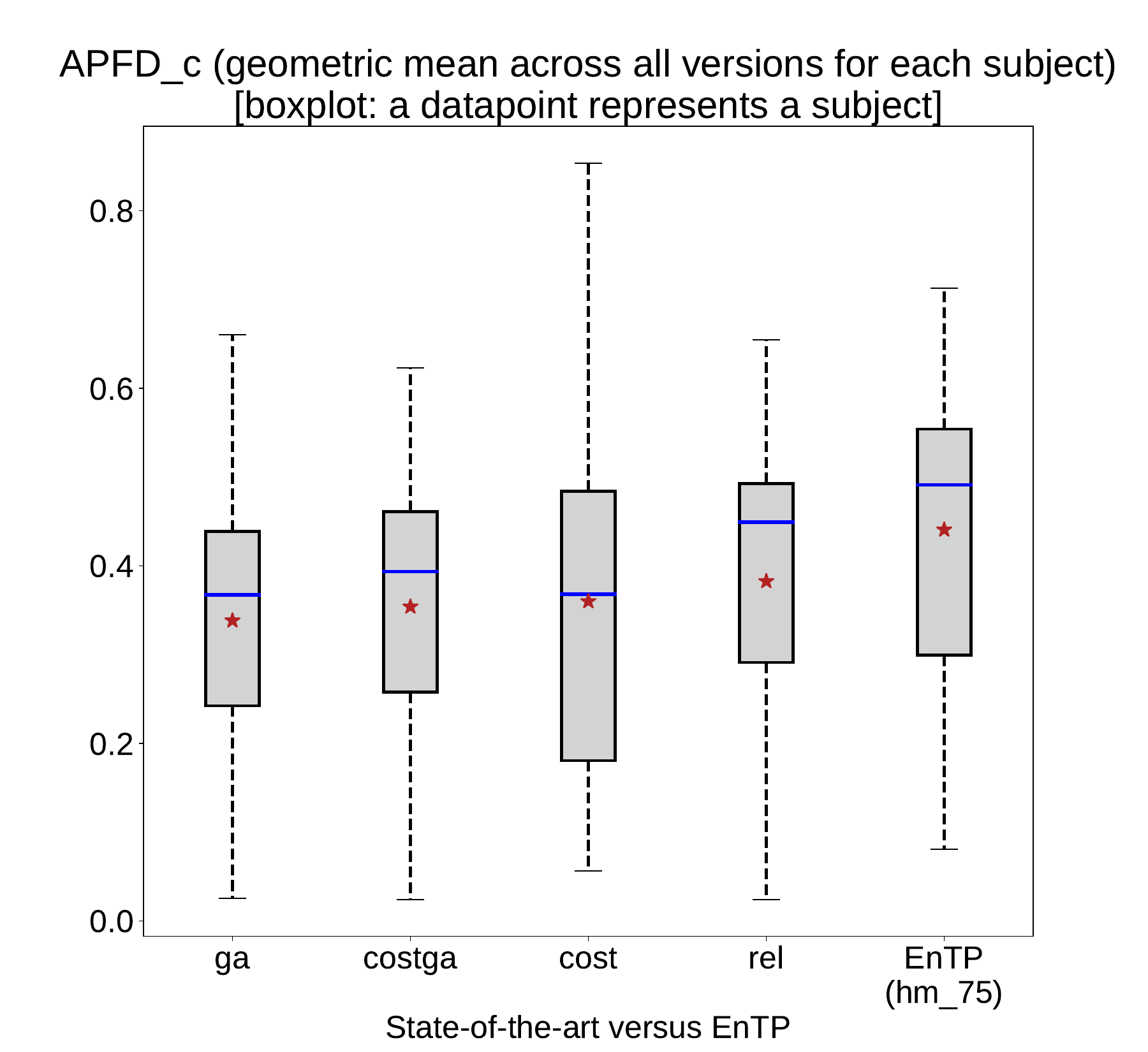}
    \includegraphics[width=0.495\linewidth]{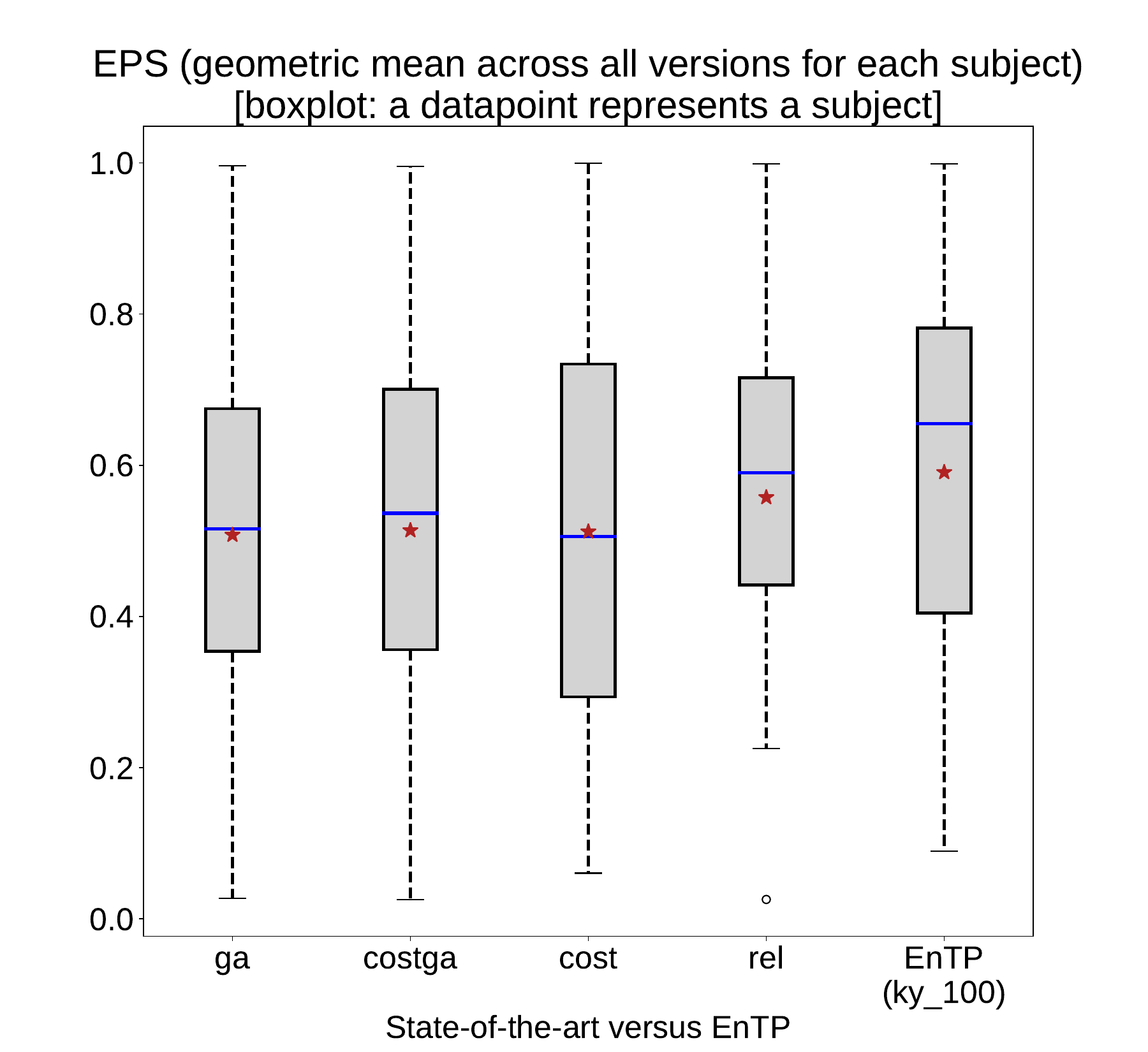}
    \caption{Effectiveness comparison against state-of-the-art prioritizations.}
    
    \label{fig:versus_SOTA}
\end{figure*}

{\fontsize{9}{9}\selectfont
	\begin{tcolorbox}
		{\textbf{\textit{\underline{Answering RQ2 (ensemble selection using diversity)}}}}
		\begin{itemize}
			\item{The efficacy of \entp comes at the diversity budget of top-75\% when Kemeny Young (KY) consensus is not considered. Irrespective of diversity budgets, KY consensus in not desirable for large test suites.}
                \item{\entp outperforms \hansie in the most realistic setting when test prioritization is performed for non-uniform test cost distributions.}
		\end{itemize}
	\end{tcolorbox}
}

\subsection{Answering RQ3 (comparison against state-of-the-art)}\label{sec:RQ3}

\noindent
\textbf{Motivation.} In this RQ, we evaluate whether \entp outperforms standalone heuristics participating in the ensemble.

\noindent
\textbf{Approach and Results.} Although \entp outperformed \hansie, it remained to be explored how state-of-the-art classical approaches compared against \entp. Figure~\ref{fig:versus_SOTA} shows that in terms of cost cognizant prioritizations effectiveness, \entp (Harmonic Mean consensus with top-75\% diversity) outperformed the four state-of-the-art approaches: \texttt{ga} (greedy additional), \texttt{costga} (cost-based \texttt{ga}), \texttt{cost} (black-box approach), \texttt{rel} (greedy total)) considered in our evaluation. For APFD ans EPS, \texttt{ky\_100} (default in \hansie, a configuration in \entp) achieved the best performance but without considering the cost of the test cases being prioritized. \entp turned out to more effective than \texttt{cost}-only approach (least effective).

{\fontsize{9}{9}\selectfont
	\begin{tcolorbox}
		{\textbf{\textit{\underline{{Answering RQ3 (comparison against state-of-the-art)}}}}}
		\begin{itemize}
			\item[]{With a diversity budget of top-75\% aggregation, \entp performs the best compared to state-of-the-art cost based prioritization heuristics in terms of cost-cognizant APFD$_c$.}
		\end{itemize}
	\end{tcolorbox}
}

\REM{
\subsection{Answering RQ4 (impact of paradoxes and ties)}\label{sec:RQ4}

\noindent
\textbf{Motivation.} The purpose of this RQ is to determine the effectiveness of the partially ordered prioritization when a perfect consensus cannot be obtained due to paradoxical scenarios and cycles in collective preferences.

\shouvick{I am here...}

\noindent
\textbf{Approach.}

\begin{figure}[h]
    \centering
    \includegraphics[width=1.04\linewidth]{images/EPS_par.pdf}
    \caption{Benefit of parallel test execution in the presence of consensus paradoxes and ties: EPS\_par values.}
    \label{fig:paradox}
\end{figure}

{\fontsize{8.5}{8.5}\selectfont
	\begin{tcolorbox}
		{\textbf{\textit{\underline{Answering RQ4 (impact of paradoxes and ties)}}}}
		\begin{itemize}
			\item[]{...}
		\end{itemize}
	\end{tcolorbox}
}

\subsection{Answering RQ5 (overhead of prioritization)}

\begin{figure}[h]
    \centering
    \includegraphics[width=1.04\linewidth]{images/ensselcomp.pdf}
    \caption{Time overhead of performing ensemble selection and rank aggregation.}
    \label{fig:ensselcomp}
\end{figure}

{\fontsize{8.5}{8.5}\selectfont
	\begin{tcolorbox}
		{\textbf{\textit{\underline{Answering RQ5 (overhead of prioritization)}}}}
		\begin{itemize}
			\item[]{...}
		\end{itemize}
	\end{tcolorbox}
}
}

\section{Threats to Validity}\label{sec:threats}

In this section, we discuss the limitations in our methodology, potential threats to validity, and their potential implications on the reliability of our experiments and conclusions.

\noindent{{\bf External validity.}}
The applicability of our approach to other experimental setups with specific benchmarking challenges may be limited. To partially address this concern, we evaluated \entp on 20 open-source programs, with 60\% of them sourced from the SIR repository and 40\% from GitHub repositories. 
The SIR sub-dataset has been widely used in empirical studies of major software testing approaches~\cite{APFD_2001,APFD_2002,Jiang_2009,Yoo_2012_survey,Assi_2018,Bertolino_2019}. In line with this practice, we used it for a fair comparison against state-of-the-art classical prioritization mechanisms that also relied on the SIR subjects. However, a potential threat remains in the unexplored behavior of \entp in generalized build systems such as \texttt{make} and \texttt{cmake}, which we plan to address in the future. To enhance generalization beyond SIR subjects, we included eight popular GitHub projects in our evaluation. These projects were previously used in a study of independent unit-test parallelization, confirming that they were free from nondeterminism and had independent test cases. Furthermore, we plan to further reduce external threats by (i) applying \entp to a broader test-bed and (ii) adapting and extending \entp to operate in the industrial setting.

\noindent{\bf Construct validity.}
The construction of the delta-displacement heuristic in our implementation of \colosseum (standalone heuristic) assumed that the overall delta-displacement would remain unchanged for each selected test case in the evolved version. However, it is possible that this approach may not be the most accurate approximation of the likelihood of fault-masking, which is closely related to failed change propagation. These potential threats need to be addressed through further empirical study. Our approach to measuring and computing may not have been the most efficient in our experiments. The assumption we made in constructing the change-displacement heuristic, that the delta-displacement would remain constant for each selected test case in the evolved version, may not accurately represent the likelihood of fault-masking, a key factor in failed change propagation. Further empirical study is needed to mitigate these threats and ensure the validity of our approach (\entp). The specific settlement of the repetition factors (4, and 7) in the expansion of the ensemble was purely based on our empirical inspection for the best result. Other better configurations may exist and we plan to mitigate this specific threat in our future studies.

\section{Related Work}\label{sec:related}
In this section, we state the position of \entp in context of existing related literature on test case prioritization.

\noindent{\bf Analysis-based approaches.} These are the most classical~\cite{Yoo_2012_survey,Kazmi_2017,Khatibsyarbini_2018,Henard_2016,Lu_2016} program analysis techniques for software regression techniques developed since the 1990s. The underlying heuristics utilize several white and black box information like source code-coverage~\cite{APFD_2002,Acharya_2011}, test-execution time~\cite{Walcott_2006}, code quality~\cite{Wang_2017_2}, risk~\cite{Huang_2014,risk_prio}, etc. to compute a test case ordering. However, \entp is orthogonal and leverages these techniques in its prepossessing step when the same set of selected regression test cases are subjected to multiple independent heuristics to get different permutations to form the ensemble of rankings. In this sense, \entp can be used in combination with the existing approaches to yield more effective prioritizations.

\noindent{\bf Learning-based approaches.} Machine Learning (ML) based approaches ~\cite{Spieker_2017_napfd,Busjaeger_2016,LTR_TCP_APSEC_2021,LTR_TCP_ICSE_2020,Chen_2018} attempt to combine the power of multiple models so as to train and build a more robust model. This is achieved by techniques broadly identified as: (i) \textit{learning-to-rank} (LTR), and (ii) \textit{ranking-to-learn} (RTL). The final permutation of test cases are obtained from this fused model either by ensemble learning, or by a \textit{non-Condorcet} voting mechanism. In contrast, \entp generates different rankings upfront due to individual heuristics (non ML-based) and then performs a social choice theoretic rank aggregation (\textit{extended-Condorcet}: Kemeny-Young, \textit{non-Condorcet}: others) which is unaware (black-box) of the code/ test case being analyzed/ executed. All previous approaches are otherwise aware (white/ gray box) and domain specific to software testing. The downside of \entp is that this may sometimes construct a non-optimal solution. However, the bright side is that biases may be removed when multiple individual heuristics have the same competing target (i.e., optimization of failure detection rate). This is achieved in \entp by the \textit{diversity based ensemble selection} and \textit{social choice theoretic consensus} computation. However, note that the term ``ensemble'' in \entp does not resemble ensemble learning. To the best of our knowledge, differing from all existing approaches in this space, our black-box ensemble aggregation approach (\entp) is not ML-based (LTR/ RTL), hence, not comparable in this aspect. However, \entp is a generic prioritization strategy and holds the power to accept as input any ranking of test cases oblivious to underlying criteria for prioritization (cost, coverage, risk, etc.).
\section{Conclusions and Future Work}\label{sec:conclu}

In this work, we discussed \entp, a new consensus prioritization approach which follows three-stage pipeline: (i) ensemble selection, (ii) rank aggregation, and (iii) test suite execution. We showed that for system level test-cases with high load imbalance within test suites, \entp shows promising performance while only aggregating top-75\% diverse rank lists (i.e., prioritizations). We envision two audiences for this work: (i) practitioners (test engineers) who can utilize \entp as a tool for regression testing, and (ii) researchers whose possible future pursuits can be made along these directions: (i) comparing \entp's efficacy at system-level with unit-level testing where test cases tend to have dependencies (verdict-wise) among themselves, (ii) considering more standalone heuristics, not only coverage based, (iii) incorporating domain knowledge (test criteria) into the diversity computation, (iv) optimizing \entp to be applicable in the Continuous Integration (CI) setting with quick release cycles.

\begin{center}
\fbox{\entp artifacts:~\url{https://doi.org/10.5281/zenodo.7278151}}   
\end{center}

\section*{Acknowledgments}
The first author is supported by the Startup Grant No. IP/IITGN/CSE/SM/2324/02 from Indian Institute of Technology Gandhinagar, India. 


\bibliographystyle{ACM-Reference-Format}
\bibliography{references}

\end{document}